\documentclass[acmsmall]{acmart}

\usepackage{packages}
\usepackage{epstopdf}
\usepackage{textpos}

\setcopyright{rightsretained}
\acmJournal{PACMNET}
\acmYear{2023} \acmVolume{1} \acmNumber{CoNEXT3} \acmArticle{24} \acmMonth{12} \acmPrice{}\acmDOI{10.1145/3629146}

\makeatletter
\gdef\@copyrightpermission{
  \begin{minipage}{0.2\columnwidth}
   \href{https://creativecommons.org/licenses/by/4.0/}{\includegraphics[width=0.90\textwidth]{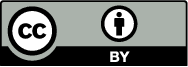}}
  \end{minipage}\hfill
  \begin{minipage}{0.8\columnwidth}
   \href{https://creativecommons.org/licenses/by/4.0/}{This work is licensed under a Creative Commons Attribution International 4.0 License.}
  \end{minipage}
  \vspace{5pt}
}

\newcommand\EatSpacesHack{\@bsphack\@esphack}
\makeatother

\iftrue
\renewcommand{\jz}[1]{\EatSpacesHack}
\renewcommand{\og}[1]{\EatSpacesHack}
\renewcommand{\ps}[1]{\EatSpacesHack}
\renewcommand{\rh}[1]{\EatSpacesHack}
\renewcommand{\mj}[1]{\EatSpacesHack}
\newcommand\reviewfix[1]{\EatSpacesHack}
\fi

\begin{document}

\setlength{\TPHorizModule}{\paperwidth}
\setlength{\TPVertModule}{\paperheight}
\begin{textblock}{0.802}(-0.025,-0.1)
  \fbox{\begin{minipage}{\linewidth}
    \noindent
    \footnotesize
    If you cite this paper, please use the following reference: Patrick Sattler, Johannes Zirngibl, Mattijs Jonker, Oliver Gasser, Georg Carle, and Ralph Holz. 2023. Packed to the Brim: Investigating the Impact of Highly Responsive Prefixes on Internet-wide Measurement Campaigns. \textit{In Proc. ACM Netw. 1, CoNEXT3, Article 24,} 21 pages. \url{https://doi.org/10.1145/3629146}
  \end{minipage}}
\end{textblock}

\title[Highly Responsive Prefixes on the Internet]{Packed to the Brim: Investigating the Impact of Highly Responsive Prefixes on Internet-wide Measurement Campaigns}

\author{Patrick Sattler}
\email{sattler@net.in.tum.de}
\orcid{0000-0001-9375-3113}
\affiliation{%
  \institution{Technical University of Munich}
  \city{Munich}
  \country{Germany}
}

\author{Johannes Zirngibl}
\email{zirngibl@net.in.tum.de}
\orcid{0000-0002-2918-016X}
\affiliation{%
  \institution{Technical University of Munich}
  \city{Munich}
  \country{Germany}
}

\author{Mattijs Jonker}
\email{m.jonker@utwente.nl}
 \orcid{0000-0001-5174-9140}
\affiliation{%
  \institution{University of Twente}
  \city{Enschede}
  \country{The Netherlands}
}

\author{Oliver Gasser}
\email{oliver.gasser@mpi-inf.mpg.de}
\orcid{0000-0002-3425-9331}
\affiliation{%
  \institution{Max Planck Institute for Informatics}
  \city{Saarbrücken}
  \country{Germany}
}

\author{Georg Carle}
\email{carle@net.in.tum.de}
\orcid{0000-0002-2347-1839}
\affiliation{%
  \institution{Technical University of Munich}
  \city{Munich}
  \country{Germany}
}

\author{Ralph Holz}
\email{ralph.holz@uni-muenster.de}
\orcid{0000-0001-9614-2377}
\affiliation{%
  \institution{University of Münster}
  \city{Münster}
  \country{Germany}
}

\renewcommand{\shortauthors}{Patrick Sattler et al.}
\begin{abstract}
    Internet-wide scans are an important tool to evaluate the deployment of
    services.  To enable large-scale application layer scans, a
    fast, stateless port scan (\eg using ZMap) is often performed ahead of time to
    collect responsive targets.  It is a common expectation that port scans on
    the entire IPv4 address space provide a relatively unbiased view as
    they cover the complete address space.
    Previous work, however, has found prefixes where all addresses share particular properties.
    In IPv6, aliased prefixes and fully responsive prefixes, \ie prefixes where all addresses are responsive, are a well-known phenomenon.
    However, there is no such in-depth analysis for prefixes with these responsiveness patterns in IPv4.

    This paper delves into the underlying factors of this phenomenon in the context of IPv4 and evaluates port scans on a total of 161 ports
    (142 TCP \& 19 UDP ports) from three different vantage points.
    To account for packet loss and other scanning artifacts, we propose the notion of a new category of prefixes, which we call highly responsive prefixes (HRPs).
    Our findings show that the share of HRPs can make up \sperc{70} of responsive addresses on selected ports.
    Regarding specific ports, we observe that CDNs contribute to the largest fraction of HRPs on TCP/80 and TCP/443, while TCP proxies emerge as the primary cause of HRPs on other ports.
    Our analysis also reveals that application layer handshakes to targets outside HRPs are, depending on the chosen service, up to three times more likely to be successful compared to handshakes with targets located in HRPs.
    To improve future scanning campaigns conducted by the research community, we make our study's data publicly available and provide a tool for detecting HRPs.
    Furthermore, we propose an approach for a more efficient, ethical, and sustainable
    application layer target selection.
    We demonstrate that our approach has the potential to reduce the number of TLS handshakes by up to \sperc{75} during an Internet-wide scan while successfully obtaining \sperc{99} of all unique certificates.

\end{abstract}

\vspace{-1em}

\ccsdesc[300]{Networks~Middle boxes / network appliances}
\ccsdesc[300]{Networks~Network measurement}
\ccsdesc[300]{Networks~Public Internet}

\keywords{highly responsive prefix, port scanning, ethical scanning}

\maketitle

\section{Introduction}
\label{sec:introduction}

In the last decade, Internet-wide port scans have been frequently used by
network and security researchers to quantify service deployment or as the first
step in a scanning pipeline that targets higher-layer protocols, \eg by
chaining ZMap~\cite{durumeric_zmap_2013} and ZGrab2~\cite{zgrab2}.
Researchers routinely assess the impact of a service
vulnerability with the help of port scans~\cite{shadowserver, shodan, divd}.
An overlooked bias can easily distort any conclusions drawn; therefore, high data quality is critical.

Previous research, \eg by \citeauthor*{izhikevich2021lzr}
\cite{izhikevich2021lzr,izhikevich2022predicting} and \citeauthor*{gasser2018clusters}
\cite{gasser2018clusters}, identified a phenomenon where all
IP addresses within a prefix seem to respond on a scanned port.
In the case of IPv6, this type of prefix is particularly important to consider.
The size of allocated IPv6 prefixes is such that a full scan is infeasible, and
hence hitlists are commonly used. If such prefixes appear on a hitlist, they are
likely to introduce an undesired bias in the scan
results~\cite{gasser2016scanning}.
Analysis of data from IPv6 scans has revealed several possible reasons for the presence of such prefixes. \reviewfix{E.4}
One of these reasons is the existence of aliased prefixes, which are prefixes where a single host responds to all addresses within that prefix~\cite{gasser2018clusters}.
Further work by \citeauthor*{zirngibl2022clusters}~\cite{zirngibl2022clusters} in 2022
found that some \ac{cdn} prefixes appear as fully responsive.  Although such
prefixes appear to be aliased, no single responder is behind an address,
either.
Therefore, the authors introduced the \ac{frp} class, which generalizes the concept of aliased prefixes in IPv6.

For IPv4, however, \acp{frp} are commonly not further considered by
researchers. In the case of \texttt{SYN} scans, scanners are fast enough to
iterate over all IPv4 addresses.
However, this does not apply to application-layer scans, which are typically
resource-heavy and take more time. Here, \acp{frp} can be much more
problematic, especially if the deployment of the targeted protocol is low or if
the same host responds for many addresses. An application-layer scan that
includes \acp{frp} will take much longer than needed and may introduce bias or
lead to unwanted artifacts in the data.
Consequently, security assessments or the evaluation of protocol deployments
can lead to incorrect conclusions.  Limiting port scans to the necessary volume
is also ethically responsible, which is an additional incentive to avoid scanning
\acp{frp}. \reviewfix{E.4}

\reviewfix{A.1, B.9}
By definition, \acp{frp} exclusively describe prefixes where every single address within a prefix exhibits responsiveness on a specific protocol-port combination.
In contrast to IPv6, where only a sample of addresses is probed to determine full responsiveness, in IPv4, we have the capability to test an entire prefix comprehensively.
Since confounding factors---such as packet loss---may impact this probing of an entire prefix,
we find the need for a new prefix classification that takes these factors into account.
In this paper, we introduce \acfp{hrp}, which are prefixes where over \sperc{90} of addresses demonstrate responsiveness for a specific port.
To the best of our knowledge, our study is the first of its kind concerning \acp{hrp} in IPv4.

\noindent\textbf{Motivating Example:}
Our interest in this subject stemmed from curiosity about the impact of observed measurement artifacts (i.e., \acp{hrp}).
When analyzing long-running measurement campaigns, we identified certain inefficiencies associated with the conventional approach of utilizing ZMap~\cite{durumeric_zmap_2013} for the detection of open ports, followed by an application layer scan on the successful addresses.

\reviewfix{C.1}
We conduct a weekly \ac{tls} scanning campaign to gather extensive \ac{tls} deployment information.
Our approach is quite typical, commencing with a preliminary ZMap scan, followed by the application-layer scan using \ac{tls}.
During this procedure, ZMap typically reports approximately \sm{50} IP addresses with TCP/443 port open.
Of these, roughly \sperc{30} are located within \acp{hrp}, with only \sperc{1.7} of \acp{as} responsible for all such \acp{hrp}.
Therefore, repeatedly scanning the same \acp{as} often fails to yield additional information, while potentially imposing a burden on infrastructure providers.
Our initial analysis showed that \acp{cdn} play a significant, albeit not an exclusive role, which prompted us to investigate more in depth.

\citeauthor{durumeric_zmap_2013}~\cite{durumeric_zmap_2013}, proposed to use ZMap in the manner mentioned---namely, as a tool for identifying responsive hosts for application-layer scans and for determining service deployment statistics.
However, our findings illustrate why this perspective may be overly simplistic and no longer entirely justified.
In \Cref{sec:related}, we provide a list of some exemplary measurement studies that have been influenced by the existence of HRPs, and evaluate the available data.

\noindent\textbf{Contributions:}
In this paper, we conduct an extensive analysis of \acp{hrp} in IPv4.
To the best of our knowledge, there has been no prior study focused on \acp{hrp} in IPv4.
We shed light on the presence and characteristics of \acp{hrp} through IPv4-based port scans from multiple vantage points.
Additionally, we present an ethical scanning approach that takes into account the existence of \acp{hrp}. \reviewfix{A.1}
In this work, we make the following contributions:

\vspace{0.2em}
\noindent
\first We conduct a comprehensive analysis of \aclp{hrp} in IPv4 across 161 ports.
Our findings reveal that \acp{hrp} consistently cover approximately \sperc{30} of responsive addresses on TCP ports 80 and 443.
Furthermore, this coverage can reach up to \sperc{70} for less common ports.

\vspace{0.2em}
\noindent
\second We evaluate \acp{hrp} from both application-layer and \ac{dns} perspectives.
Our analysis shows that targets outside \acp{hrp} exhibit up to three times higher responsiveness at the application layer compared to targets inside \acp{hrp}.

\vspace{0.2em}
\noindent
\third We provide a tool for detecting \acp{hrp} based on port scan output, such as that generated by ZMap.
Additionally, we publish detailed statistics for all ports analyzed~\cite{datatum} and offer continuous weekly \ac{hrp} results for TCP ports 80 and 443 on our website -- \textbf{\url{https://hrp-stats.github.io/}}.

\vspace{0.2em}
\noindent
\fourth We discuss our findings and the potential for more ethical scanning practices in the future.
Our proposed adjusted scanning approach enhances scan success rates by selectively targeting specific hosts within \acp{hrp}, thereby reducing the load on scanned infrastructure.
Our results demonstrate that, with this approach, we can decrease the number of application-layer scans by up to \sperc{70} while retaining all essential information about the target prefixes.

\noindent\textbf{Outline:}
Related work is covered in \Cref{sec:related}.
In \Cref{sec:datasets}, we introduce the conducted scans and datasets we used.
In \Cref{sec:methodology}, we define \aclp{hrp} and explain our approach and tooling to detect them.
We evaluate the presence, stability, and specifics of \acp{hrp} in \Cref{sec:evaluation}.
Finally, we discuss our proposed scanning approach with advantages and disadvantages in \Cref{sec:discussion}.
We discuss ethical considerations in \Cref{sec:ethics} and conclude our paper in \Cref{sec:conclusion}.
\section{Related Work}
\label{sec:related}

\noindent\textbf{IPv4 Tarpits:}
Tools such as ZMap \cite{durumeric_zmap_2013} and MASSCAN \cite{masscan} made it feasible to conduct Internet-wide port scans.
However, it is crucial to exercise caution when analyzing their output results, particularly in terms of the reachable services they may have identified. \reviewfix{D.7}
In 2014, \citeauthor*{alt2014v4aliases} \cite{alt2014v4aliases} analyzed
so-called tarpits, cyber-defense tools which imitate many fake hosts to slow
down scanners.
As tarpits are often deployed on a whole prefix, they look like \acp{hrp} on the Internet.
Moreover, tarpits behave like responsive hosts but do not offer any real service.
Therefore, our evaluation of \ac{hrp} application-layer responsiveness shows that tarpits play only a minor role in our dataset.

We do not focus on this single reason for \acp{hrp}, but quantify their
occurrence and specifics in general.  \sm{1.9} IPv4 addresses were also assumed
to be tarpits by \citeauthor*{bano2018liveness} \cite{bano2018liveness}.  Their
argument rested on the responsiveness on a high port as such a port is unlikely to be responsive on many consecutive addresses.
Similarly, we show that some \acp{hrp} are responsive on all ports, including
high ports unrelated to any service.  In 2021, \citeauthor*{izhikevich2021lzr}
\cite{izhikevich2021lzr} investigated the phenomenon of hosts responding to
\texttt{SYN} scans but not following through with the application-layer
handshake performing probes to \sperc{0.1} of the IPv4 address space.
The authors built a tool to identify services running on unexpected ports,
showing that many common services are in practice deployed on ports other than
the default or well-known ones.
They identified full IPv4 blocks larger than \texttt{/24} with
a zero TCP initial window but did not further investigate on prefix level
responsiveness.
In \Cref{sec:evaluation}, we analyze the presence of \acp{hrp} in the dataset of \citeauthor*{izhikevich2021lzr} and validate our claims using their data.

\noindent\textbf{Predicting Responsiveness:}
A more recent study by \citeauthor*{izhikevich2022predicting} \cite{izhikevich2022predicting} proposed an approach to predict service availability on non-standardized alternative ports.
They perform small scale scans in order to deduce availability of addresses on the Internet.
In this work, we show that the alternative and non-standardized ports suffer the most bias through \acp{hrp}.
Although \citeauthor*{izhikevich2022predicting} verify service availability using application layer handshakes, we argue that considering our findings can improve the success rate and performance of such measurement campaigns.

\reviewfix{B.2}
Additional work by \citeauthor{durumeric_censys_2015}~\cite{durumeric_censys_2015} introduced the Censys search engine.
Our findings also extend to search engines such as Censys and the data they provide to researchers.
ZMap has been employed for various purposes, including determining the number of vulnerable Heartbleed hosts~\cite{durumeric_heartbleed_2014} and tracking patching progress.
Furthermore, \citeauthor{costin2014embedded}~\cite{costin2014embedded} employed ZMap in the intended manner, identifying hosts with open TCP/443 ports for conducting HTTPS probes.
Throughout this paper, we will demonstrate why it has become increasingly crucial to account for the presence of \acp{hrp} when conducting such evaluations.

\noindent\textbf{Ethical Scanning:}
\citeauthor*{klick2016tass} \cite{klick2016tass} introduced a scanning approach
that reduces probing overhead by focusing on interesting, densely populated
prefixes based on historical data.  We confirm that \acp{hrp} are often stable
and large fractions of responsive addresses are within \acp{hrp}.  However,
focusing on those potentially introduces biases in scans towards \acp{cdn} and TCP proxies.
Moreover, we show that the application-layer success rate is higher outside
\acp{hrp}, their approach would potentially omit these non-\ac{hrp} targets.

\noindent\textbf{IPv6 Aliased Prefixes:}
Subtypes of \acp{hrp} have been regularly discussed in IPv6 research.
During IPv6 scans, even a single \ac{hrp} can already be too large to be practically scanned.
\citeauthor*{murdock20176Gen}~\cite{murdock20176Gen},
\citeauthor*{gasser2018clusters}~\cite{gasser2018clusters}, and others \cite{luckie2013speedtrap, padmanabhan2015uav6,beverly2013tbt} identified so-called
aliased prefixes that impact the statistical evaluation of scans.  They
attributed the effect to a single host that aliased an entire prefix.  The authors
developed means to detect this based on randomly sampled addresses within each
prefix.  \citeauthor*{gasser2018clusters}~\cite{gasser2018clusters} included
such a detection in their ongoing hitlist service.  However, in 2022,
\citeauthor*{zirngibl2022clusters}~\cite{zirngibl2022clusters} showed that
aliased prefixes are not necessarily a single host, \eg
in \acp{cdn}.
We find \acp{hrp} announced by similar \acp{as} in IPv4, and
show their visibility for ports not investigated by
\citeauthor*{zirngibl2022clusters}.
\section{Datasets}
\label{sec:datasets}

In this section, we describe the data sources we rely on for our analysis later in
\Cref{sec:evaluation}.  Our sources specifically include port, HTTPS and
\ac{dns} scans that we conduct ourselves, as well as (semi)public data sources
that we use as complement.
We discuss ethical considerations for our scans in \Cref{sec:ethics}.
The datasets we obtain from the various sources mainly cover
the first week of August 2022. We will note where this does not
apply.

\subsection{Port Scans}
We conducted different port scans from three vantage points.
Two vantage points (\muc and \scn) are located in Germany, while the third is
located in Australia (\syd).
The first two vantage points are used for most of the scans.
The third was used in sync with \muc to
evaluate the visibility of \acp{hrp} from two geographically and topologically
distant vantage points (see \Cref{ssec:bias}).
\reviewfix{B.1}
We use ZMap and its TCP SYN scan module to perform these scans.
The used scan module reports a target as being responsive, if a TCP SYN-ACK packet has been received from that target.

We run two types of scans from \muc and \scn:

\noindent
\first Long-running scan campaign for TCP port 443 on the announced IPv4 address space (Jan 2021 --- Jan 2023). We use the resulting data to analyze the
stability of \acp{hrp} over time.

\noindent
\second A wider-range IPv4 scanning campaign targeting 36 different TCP ports (Aug 2022
-- Sep 2022).  We complement this scan data using port scan data
from Rapid7's Project Sonar~\cite{rapid7}. Our scans allow us to evaluate \acp{hrp} while considering
potential artifacts on our scan machines, of the used blocklist, and scanned
targets.

\subsection{HTTPS Scans}
Besides port scans, we run weekly HTTPS scans on responsive hosts obtained through the port scans.
These application layer scans are conducted from \muc using the Goscanner \cite{goscanner}, with and without \ac{sni}.
The weekly TCP/443 port scans are used as the target list for the scan without \ac{sni}.
The \ac{sni} scan uses a locally resolved \texttt{A} records dataset for more than \sm{350} domain names.

\subsection{DNS Data}
We use DNS data provided by the OpenINTEL project, which actively queries a significant part of all globally registered domain names on a daily basis~\cite{openintel}.
The OpenINTEL measurement is primarily seeded using zone files, including well over \sk{1} zone files from ICANN's CZDS~\cite{icannczds} as well as country-code top-level domain zones.
The measurement data include address resolutions for domain names as well as for infrastructure records (i.e., MX and NS).
The associated A records allows us to map domain hosting and mail exchangers and authoritative name servers to HRPs.

\subsection{Rapid7 Data}
To complement our own scans, we use data provided by Rapid7's Project Sonar~\cite{rapid7}.
\reviewfix{D.2} Rapid7 provides measurement data from Internet-wide surveys to gain insights into global exposure to common vulnerabilities containing regularly-conducted port scans on
129 different TCP ports (\eg 80, 443, 25, 60000) and 19 UDP ports, (\eg port 53). Note that common ports (\eg 80
or 443) are scanned more frequently by Rapid7. We use these frequent scans to determine the stability  of \acp{hrp} besides our results.
Rapid7 also provides application-layer scan data.
We rely on Rapid7 data to reduce the need to run our own measurements
and to reach a large coverage of different ports and protocols.

We use different datasets provided by Rapid7:

\noindent
\first
For our analysis, we use port scans conducted by Rapid7 from February 1, 2021 to the end of 2022.

\noindent
\second We use \ac{tls} scans to evaluate \acp{hrp} in more
detail. Rapid7 performs scans for TLS ports other than TCP/443, so we
do not have to run these ourselves.  We use the TLS handshake status
and the collected certificate from scans in
the first week of August 2022.

\noindent
\third We also use application-layer HTTP scans for port 80 and alternative
HTTP ports from Rapid7. These contain the complete raw HTTP responses received by Rapid7.
\section{HRP Detection Approach}
\label{sec:methodology}

This section details how we collect, aggregate, and enrich reachability data per prefix and finally identify a responsiveness threshold for \acp{hrp}.

\subsection{Data Collection and Aggregation}
Our tool takes as input a list of IPv4 addresses that are reachable on a specific port, \eg from a ZMap scan.
Therefore, \acp{hrp} are specific to a protocol and a corresponding port.
In \Cref{ssec:port_eval}, we evaluate cross-port responsiveness.
As a first step, we aggregate these input addresses into prefixes.
A prefix length of 24 bits is a de facto limit for the most-specific, globally routable prefix length in IPv4
\cite{sediqi2022hyper,rfc7454,MANRS2021Filter}.
Thus, this constitutes the smallest IPv4 prefix that reliably propagates in the \ac{bgp}, representing the most limited subset of addresses we evaluate.
\Cref{fig:reachable_addrs_per_pfx} shows that more-specific prefix sizes do not amount to a substantial share of high responsiveness in our analysis. \reviewfix{E.1}
In \Cref{ssec:port_eval}, we show that our results indeed support this choice.

When aggregating responsive addresses into \texttt{/24} prefixes, we
enrich each entry with additional contextual information, such as the origin AS
number and the corresponding \ac{bgp} prefix, using \ac{bgp} dumps from a Route
Views~\cite{routeviews} collector of the respective scan date.
Our approach is as efficient as the aggregation efficiency as we read the input
data only once and linearly after aggregating it.  This processing strategy enables parallelized
(\eg sharded) processing.

\subsection{Identifying a Threshold for Highly Responsive Prefixes}
\label{ssec:hrp_definition}
With the prefix statistics at hand, we next need to identify a suitable
threshold for \acp{hrp}.  Previous work mentioned the existence of fully
responsive prefixes in IPv4~\cite{izhikevich2021lzr}.
However, to the best of our
knowledge, there has been no attempt yet to identify the various categories that
together constitute \acp{hrp}.
The need for a generalized definition that covers prefixes with similar behaviors arises from several aspects that bear relevance for scanning:
\first Packet loss can occur, causing a prefix to appear as not \sperc{100} responsive, even when it actually is.
It is important to note that even a minimal loss rate can substantially impact an Internet-wide IPv4 scan.
\second
Scalable port scanners such as ZMap sometimes trigger rate limiting on
the receiver side.
While built-in target randomization helps mitigate the risk of
overburdening target systems, it cannot completely eliminate this problem.
We also identify some instances of rate limiting and evaluate them in \Cref{ssec:bias}.

\begin{figure}
	\centering
	\begin{subfigure}{0.45\textwidth}
		\includegraphics{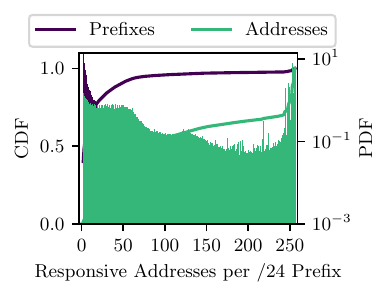}
		\caption{TCP/443}
		\label{fig:reachable_addrs_443_per_pfx}
	\end{subfigure}
	\hfill
	\begin{subfigure}{0.45\textwidth}
		\includegraphics{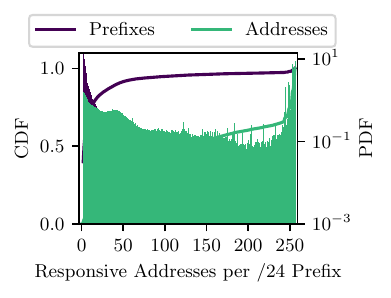}
		\caption{TCP/80}
		\label{fig:reachable_addrs_80_per_pfx}
	\end{subfigure}

	\caption{\reviewfix{B.9} Respective (bars -- right y-axis) and cumulative (line -- left y-axis) probability distribution of responsive addresses inside prefixes. The address data represents the influence of prefixes on the scan results.The prefix indicates that the majority of prefixes do not meet the criteria for \acp{hrp}. Note the logarithmic axis for the right Y-axis. More ports can be found on our published website~\cite{hrpwebsite} and in the appendix \Cref{fig:reachable_addrs_8443_8080_per_pfx}.}
	\label{fig:reachable_addrs_443_80_per_pfx}
\end{figure}

\begin{figure}
    \centering
    \includegraphics{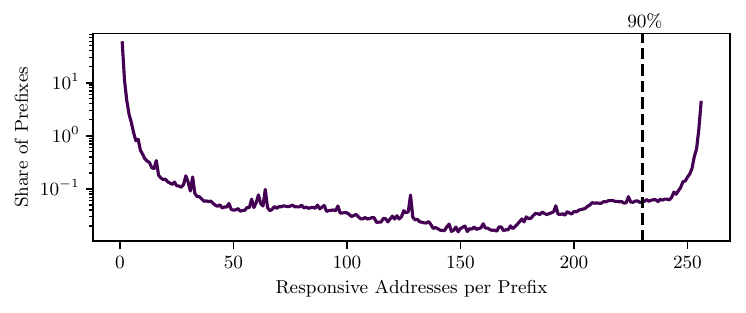}
    \caption{Average probability distribution of responsive addresses inside prefixes for all prefixes. Note the logarithmic Y-axis. The vertical line marks the \sperc{90} threshold.}
    \label{fig:reachable_addrs_per_pfx}
\end{figure}

To obtain a viable and robust responsiveness threshold, we analyze the
distribution of responsive addresses inside a prefix.  In
\Cref{fig:reachable_addrs_443_per_pfx}, we show a very relevant pattern for
TCP/443 scans. More than \sperc{75} of prefixes contain fewer than \num{15}
responsive addresses, and \sperc{91} include at most \num{50}
addresses.  However, there is a substantial increase in prefixes with \num{231}
or more responsive addresses (\ie \sperc{90} of a \texttt{/24} prefix).
Although these account for only \sperc{2.2} of \emph{visible \texttt{/24}
prefixes}, they account for \textit{\sperc{30} of responsive addresses}.  A
similar pattern is visible for TCP/80 in \Cref{fig:reachable_addrs_80_per_pfx}.
A comprehensive analysis encompassing all evaluated ports (see \Cref{fig:reachable_addrs_per_pfx}) reveals a consistent pattern among all of them.
Furthermore, separate evaluations on other ports are available in the appendix (see \Cref{fig:reachable_addrs_8443_8080_per_pfx}) and on our website~\cite{hrpwebsite}.
These supplementary evaluations also exhibit a similar pattern to what is observed for TCP/80, TCP/443, and the aggregated analysis.
Consequently, this observation holds true not only for specific ports but also extends to all other ports.

To ascertain enough generality, we inspect the curve's knee and identify
\sperc{90} of responsive addresses as a suitable threshold. In
\Cref{sec:evaluation}, we compare our choice to a \sperc{95} threshold and show
that this leads to only marginal differences in the outcome of our
investigation.
\section{Evaluation of Port Scan HRPs}
\label{sec:evaluation}

In this section, we evaluate the presence, stability, and origin of \acp{hrp}.
Furthermore, we analyze whether application-layer services actually run in these prefixes.
We use these insights to identify different classes of \acp{hrp}.

\subsection{Comparison of Ports and Protocols}
\label{ssec:port_eval}

We apply our definition of \acp{hrp} from \Cref{ssec:hrp_definition} to all available port scan data to obtain a more complete overview. In total, we analyzed scan data for \num{142} TCP and \num{19} UDP ports.
Due to space constraints, we report only on well-known or particularly interesting ports here. Evaluations for all ports are available on our website~\cite{hrpwebsite}.

\begin{figure}
    \begin{subfigure}{0.45\textwidth}
        \includegraphics{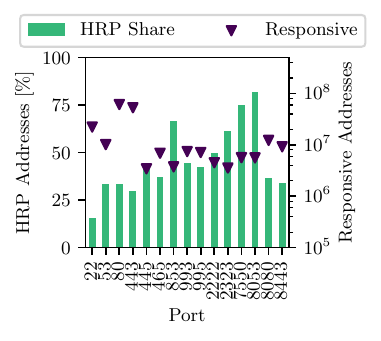}
        \caption{Selected TCP ports from our scan.}
        \label{fig:local_port_comparison}
    \end{subfigure}
    \hfill
    \begin{subfigure}{0.45\textwidth}
        \includegraphics{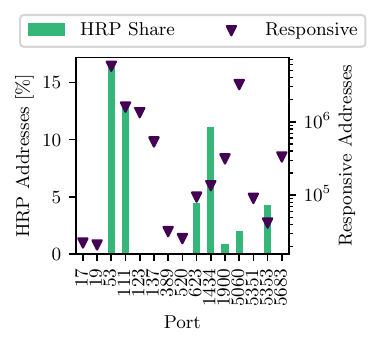}
        \caption{Rapid7 UDP results.}
        \label{fig:udp_port_comparison}
    \end{subfigure}
    \caption{\ac{hrp} address share and total number of responsive addresses per port. Note the logarithmic right Y-axis and the different Y-axis scaling for TCP and UDP.}
    \label{fig:proto_port_comparison}
\end{figure}

\begin{figure*}
    \centering
    \begin{subfigure}{0.45\textwidth}
        \includegraphics[width=\textwidth]{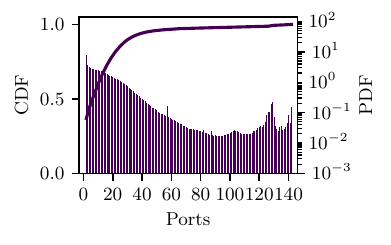}
        \caption{All prefixes.}
        \label{fig:all_pfxes_ports}
    \end{subfigure}
    \hfill
    \begin{subfigure}{0.45\textwidth}
        \includegraphics[width=\textwidth]{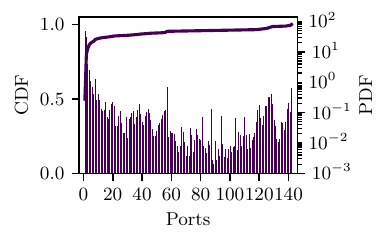}
        \caption{Highly reachable prefixes only.}
        \label{fig:hrp_ports}
    \end{subfigure}
    \caption{Number of ports on which addresses within a \texttt{/24} prefix respond. The \ac{hrp} analysis counts only ports which appear to be highly responsive for the prefix. Note the logarithmic right Y-axis.}
    \label{fig:pfx_port_comparison}
\end{figure*}

\Cref{fig:local_port_comparison} shows the share of addresses in \acp{hrp}, measured against the full set of IP addresses that respond on a given port. The figure shows data from our scans (see \Cref{sec:datasets}).
Except for the SSH port (TCP/22), at least \sperc{30} of the responsive addresses are located within \acp{hrp}. The SSH port is considered sensitive, which may be why operational configurations treat it differently.
Well-known ports (\eg TCP/80, TCP/443) and their alternative ports (TCP/8080 and TCP/8443) have an \ac{hrp} address share of \sperc{30}-\sperc{40}.
Ports for less popular services and other alternative ports generally have a higher share of \ac{hrp} addresses, which goes up to \sperc{80}.
In general, the smaller the overall set of responsive addresses and the less known a port is, the higher the fraction of \ac{hrp} addresses will be.
One reason for this is a group of \acp{hrp} that are responsive on all scanned ports (see \Cref{fig:pfx_port_comparison}).

\Cref{fig:udp_port_comparison} shows our results for UDP ports, taken from Rapid7 scan data.
Compared to the TCP scans, the absolute responsiveness is drastically smaller by at least one order of magnitude.
UDP scans generally require valid payloads to determine whether a port is truly responsive.
However, the fraction of addresses within \acp{hrp} is smaller. We still find some \acp{hrp}, especially for more well known ports, \eg 53.

To get further insights into the specifics of \acp{hrp}, we analyze whether they are responsive for a single port, specific port combinations, or all ports in our dataset.
\Cref{fig:all_pfxes_ports} shows the number of ports where at least one address is responsive inside the prefix.
The same but limited to \acp{hrp} is plotted in \Cref{fig:hrp_ports}.
The peak at 129 ports in both figures is because Rapid7 scans only 129 TCP ports, and we merged our port scanning set with Rapid7.
Some \acp{hrp} appear not responsive in our scans (see \Cref{ssec:bias}).
Moreover, fewer prefixes are highly responsive on a large share of scanned ports in Rapid7 data.
While our scans find \sperc{31} of detected \acp{hrp} being classified as highly responsive for ten or more ports, Rapid7's data only accumulates to \sperc{13.6} (our scans also accumulate in absolute numbers to more than double the Rapid7's amount).

We note the more pronounced values for \acp{hrp}: \sperc{80} appear only for three ports or fewer.
\sperc{50} of all distinct \acp{hrp} are only classified as such based on the responsiveness of a single port.
For port TCP/80, we find \sk{59.8} prefixes; for port TCP/443, we find \sk{40.7}; and for port TCP/25, we find \sk{33.8}.
\sperc{30} of prefixes are classified as \ac{hrp} based on two responsive ports 80, and 443.
However, \num{4700} (\sperc{1}) \acp{hrp} are responsive on at least 139 ports within our data.
%
%
%
%
%
%
%

\noindent
\textit{\textbf{Key take-away:}
We find \acp{hrp} for all evaluated TCP ports and show that they cover between \sperc{30} and up to \sperc{80} of all addresses responsive on a given port.
While \sperc{80} of \acp{hrp} are classified as such based on one or two responsive ports, \sperc{5} are responsive on more than 120 ports.
Due to the nature of UDP scanning, the role of \acp{hrp} is less prominent. However, we can still find instances of UDP \acp{hrp}.
}

\subsection{Stability over Time}
\label{ssec:stability}

In this section, we analyze our long-running scans on TCP/443 together with scans of TCP/80 by Rapid7 to investigate the stability of \acp{hrp} over time. Apart from understanding the long-term behavior of \acp{hrp}, this also provides evidence that the analyses of the data collected for this paper do not depend on the time a scan was carried out.

\begin{figure}
	\centering
	\includegraphics[width=\linewidth]{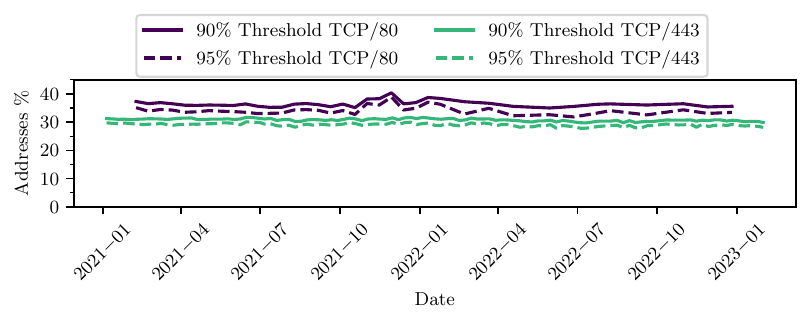}
    \caption{Stability of \aclp{hrp} for weekly scans.}
	\label{fig:address_stability}
\end{figure}

\Cref{fig:address_stability} shows the fraction of addresses within \acp{hrp} for each weekly port scan from \muc.
We also compare our \ac{hrp} definition, which is based on \sperc{90} responsiveness, with an alternative definition based on \sperc{95} responsiveness. We show that the difference is minimal, even over time.
The data from Rapid7 shows slightly more variance, but the difference is at most \sperc{5}.
These numbers imply no significant increase in address responsiveness within each \ac{hrp}.
We obtained similar findings for other services and alternative HTTPS ports.

Turning our attention to the stability of individual \acp{hrp}, we observe that most prefixes exhibit a high degree of stability, although some churn is evident.
During each TCP port 443 scan, an average of \sk{65.4} \acp{hrp} are identified, amounting to \sk{110} distinct \acp{hrp} over the two-year observation period.
\sk{78} of these remain consistently visible and classified as \acp{hrp} for at least half of our observation period, and \sperc{38.3} of these maintain their classification throughout the entire observation period.
\sperc{73.1} of prefixes that are normally classified as \acp{hrp} are missing this classification for at most five scans.
It is plausible that these misclassifications are due to scanning artifacts, \eg because of rate limiting on the receiver's side (see \Cref{ssec:bias}).
The prevalence of \acp{hrp} that we have identified, both spatially and over time, in this and the preceding section underscores the importance for researchers to account for their influence when analyzing scan data.
\reviewfix{D.7}

\noindent
\textit{\textbf{Key take-away:}
The contribution of \acp{hrp} to the overall share of responsive addresses remains consistent over a span of two years.
Over a year, more than \sperc{73} of \aclp{hrp} remain consistently detectable in our scans.
Consequently, it is crucial to account for \acp{hrp} in the analysis of both historical and recently collected scan data.
}

\subsection{Spatial Stability}
\label{ssec:bias}

\begin{figure}
    \includegraphics{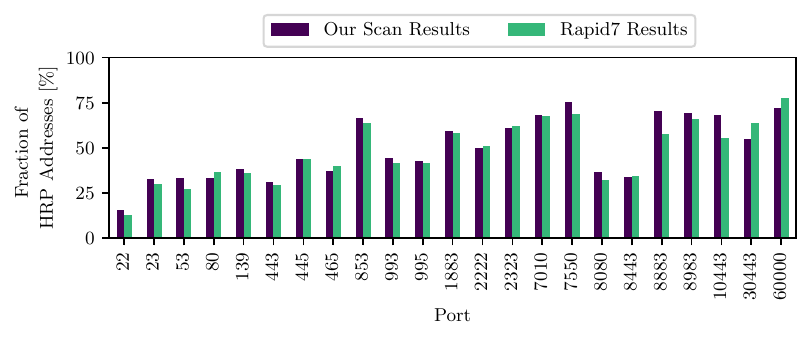}
    \caption{Comparing \ac{hrp} shares of Rapid7 data and our own scanning results.}
    \label{fig:port_comparison}
\end{figure}

In order to determine the influence of the vantage point on the detection of \acp{hrp}, we
run different measurements and comparisons between geographically and topologically diverse vantage points.
\Cref{fig:port_comparison} compares our port scans from August 2022 with Rapid7's results for 23 ports from the same time frame.
While we have observed minor variations, there are no significant discrepancies between our results and those obtained by Rapid7.

To gain deeper insights into why such discrepancies arise, we conducted parallel scans from \muc (Europe) and \syd (Oceania) employing the same ZMap parameters.
This approach ensures a fully comparable scan result as all hosts are scanned at about the same time and from different vantage points.
We use TCP/443 and TCP/55974 as the destination ports, the latter being a random ephemeral port to find differences in \ac{hrp} deployments on all ports.

We have determined that \sperc{1} of \acp{hrp} identified at one vantage point are not classified as \acp{hrp} at another.
The majority of these unclassified prefixes have no responsive hosts.
Upon conducting additional verification scans from \muc on prefixes that were exclusively detected by \syd, we found that \sperc{18} of these prefixes displayed responsiveness in the second scan and were accurately categorized as \ac{hrp}. \reviewfix{D.7}
The responsive prefixes are part of the Cloudflare \ac{as} (AS13335).
\syd has a similar number of Cloudflare prefixes not visible in its scan but rated as \ac{hrp} by \muc.
Therefore, we conclude this to be an artifact of rate limiting on Cloudflare's side.
The remaining prefixes, which are only seen by \syd, are still unresponsive to the \muc scan.
We assume this to be blocking of scans from \muc, possibly due to previous, long-running measurement campaigns from the same \texttt{/24} subnet.

TCP/55974 results show a similar number of differences with three \acp{as} (DoD: AS721 and AS5972, Illinois Century Network: AS6325) contributing \sperc{96} of the missing \acp{hrp} in the \muc results.
As DoD accounts for \sperc{11} of visible prefixes and \sperc{39} of responsive hosts, we also attribute this effect to rate limiting.

\reviewfix{B.5}
Similar insights on the impact of vantage points have been reported by \citeauthor*{wan2020originofscanning}~\cite{wan2020originofscanning} in 2020.
Given that Rapid7 may not be utilizing precisely identical parameters, conducting scans from a distinct network (one that hosts several Internet-wide scanning campaigns), and is scanning hosts at different points in time, the discrepancies between our results and those of Rapid7 are minimal.
These variations are expected and align with the outcomes of our controlled experiments.

\noindent
\textit{\textbf{Key take-away:}
We find differences in the order of \sperc{1} between different vantage points when applying the same scanning procedure.
These results confirm previous results by \citeauthor*{wan2020originofscanning}~\cite{wan2020originofscanning}.
Well-known ports have comparably fewer differences, primarily due to their higher responsiveness.
The majority of discrepancies can be attributed to rate limiting.
Consequently, we conclude that it is feasible to utilize detected \ac{hrp} prefixes from different vantage points.
}

\subsection{Origin \acp{as}}
\label{ssec:as_analysis}

The \acp{hrp} we find are announced by \sk{67.2} origin \acp{as}.
We determine the origin \ac{as} with BGP dumps from RouteViews~\cite{routeviews} from the day of the respective scan.
Next, we investigate which \acp{as} originate \acp{hrp} and to what degree an \ac{as} may deploy features that make a prefix an \ac{hrp}.

Of the \acp{as} that originate \acp{hrp}, \sperc{42} have \acp{hrp} for a single port; \sperc{76} of \acp{as} have \acp{hrp} with at most five different, responsive ports.
118 \acp{as} announce at least one \ac{hrp} where the IP addresses respond on all ports.
\Cref{tbl:ases_addresses} shows the top ten \acp{as} with at least one \ac{hrp}.
Cloudflare (AS13335), for example, announces a /16 that is fully responsive on all ports.
\reviewfix{B.7}
We identified that Spectrum~\cite{cfspectrumblog, cfspectrumports, cfspectrum}, a reverse TCP/UDP proxy for DDoS protection, is related to the affected range.
We contacted Cloudflare and confirmed with them that the service is built on the principle of binding to all ports.
Google, the second largest \ac{as} with an \ac{hrp} for all ports, offers a similar service.

\begin{table}
	\centering
	\footnotesize
	\caption{Top 10 \acp{as} based on the number of \acp{hrp} we detect across all scans in August 2022. The \ac{hrp} share is the degree to which an \ac{as} is filled with \acp{hrp}. \reviewfix{B.6}}
	\label{tbl:ases_addresses}
	\begin{tabular}{llS[table-format=3.1, table-space-text-post = \si{\kn}]S[table-format=2.1, table-space-text-post = \si{\kn}]S[table-format=2.1, table-space-text-post = \si{\percent}]SS}
		\toprule
		\ac{as} & & {Visible /24} & {\acp{hrp}} & {\ac{hrp} Share} & {Visible Ports} & {Ports with \acp{hrp}} \\
		\midrule
		Akamai & AS16625 & 22.9 \si{\kn} & 22.4 \si{\kn} & 97.8 \si{\percent} & 5 & 3 \\
		Akamai & AS20940 & 24.7 \si{\kn} & 21.1 \si{\kn} & 85.6 \si{\percent} & 136 & 5 \\
		Telin & AS7713 & 12.5 \si{\kn} & 6.5 \si{\kn} & 52.5 \si{\percent} & 136 & 4 \\
		Amazon & AS16509 & 134.9 \si{\kn} & 6.0 \si{\kn} & 4.4 \si{\percent} & 136 & 135 \\
		DoD & AS721 & 4.9 \si{\kn} & 4.5 \si{\kn} & 91.3 \si{\percent} & 136 & 55 \\
		DoD & AS5972 & 4.5 \si{\kn} & 4.4 \si{\kn} & 99.3 \si{\percent} & 60 & 54 \\
		du & AS15802 & 4.3 \si{\kn} & 4.1 \si{\kn} & 96.0 \si{\percent} & 136 & 3 \\
		Cloudflare & AS13335 & 3.1 \si{\kn} & 3.0 \si{\kn} & 98.3 \si{\percent} & 136 & 136 \\
		Cogent & AS174 & 17.0 \si{\kn} & 2.8 \si{\kn} & 16.5 \si{\percent} & 136 & 136 \\
		TANet & AS1659 & 8.2 \si{\kn} & 2.4 \si{\kn} & 29.8 \si{\percent} & 136 & 135 \\
		\bottomrule
	\end{tabular}
\end{table}

In \Cref{tbl:ases_addresses}, we find Akamai at the top of the list with \sk{22} \acp{hrp} in AS16625 and \sk{21} in AS20940.
The latter AS has nearly all of its prefixes where we find responsive addresses also classified as  \acp{hrp}.
Notable other \acp{as} are Amazon, which has \sk{6} \acp{hrp}, but these cover only \sperc{4.4} of its reachable prefixes.
The \acp{as} with an \ac{hrp} on all ports are all large network infrastructure providers: Cloudflare, Amazon, Google, Cogent, and TANet\footnote{Amazon and TANet are only missing out on a single port. See \Cref{ssec:bias} for more information.}.
It is not too surprising that \acp{cdn} are commonly connected to \acp{hrp}: they deploy various techniques~\cite{fayed2021addressagility} to support millions of domains served on limited address space.
\reviewfix{B.6} \acp{as} primarily providing \ac{cdn} services consist of up to \sperc{99} of \acp{hrp} compared to their overall announced address space (see \ac{hrp} share in \Cref{tbl:ases_addresses}).
However, this impact on scan data has so far received little attention.

We also evaluate the \ac{as} distribution per port.
For port TCP/443, Akamai has the most \acp{hrp}, followed by Google, Cloudflare, Amazon, and Fastly.
These five organizations cover \sperc{64} of all port 443 \acp{hrp}.
While Akamai's AS20940 has more port 443 \acp{hrp}, AS16625 (also Akamai) has more than double the amount of \acp{hrp} on port 80 compared to port 443.
This result may reflect different use cases that the prefixes inside these \acp{as} serve.

\noindent\textbf{CDNs affinity to HRPs:}
\reviewfix{C.2}
We identified several \ac{cdn} \acp{as} in our results that are nearly entirely populated with \acp{hrp}.
For instance, Akamai exhibits \sperc{97.8} and \sperc{85.6} of announced /24 prefixes classified as highly responsive for two different ASes, while Cloudflare demonstrates \sperc{98.3}, and Fastly registers \sperc{87.6}.
Therefore, we set out to uncover the reasons for \acp{cdn} using \acp{hrp}.
For Cloudflare, we find a paper from \citeauthor{fayed2021addressagility}~\cite{fayed2021addressagility} with an accompanying blog post~\cite{cfaddressagility} that explains Cloudflare's \textit{addressing agility} approach.
This technique decouples IP addresses from domain names and services.
The authoritative name server can select the addresses in the query response from a full prefix.
Hence, all addresses inside this prefix have to be responsive, and Cloudflare's approach needs to handle all prefix-assigned services.
We verified this assumption by registering a test website with Cloudflare, resolving the domain to an A record, and connecting to several addresses inside the A record's /24 prefix.
Each TLS handshake, which included our registered domain name in the server name indication extension to any address inside this prefix, resulted in the same certificate for our domain name.

\noindent
\textit{\textbf{Key take-away:}
    In this section we show the that the distribution across \acp{as} is dominated by \acp{cdn} and other content providing \acp{as}.
	\acp{hrp} make up a substantial part of these \acp{as}.
	We could confirm that Cloudflare's address agility technique is responsible for Cloudflare's large share of \acp{hrp}.
	We assume that other CDNs deploy similar techniques.
}

\subsection{Application-Layer Results}
\label{ssec:app_layer_results}

A number of published papers \cite{izhikevich2021lzr,zirngibl2022clusters,bano2018liveness} have reported that an open port does not necessarily imply that the corresponding application is going to respond as well.
We use application-layer scans to reveal whether the services with open ports in the respective \acp{hrp} actually respond.
\citeauthor*{izhikevich2021lzr}~\cite{izhikevich2021lzr} showed that the services running on given ports can vary widely, especially for ports that are not among the well-known ones.
Therefore, our analysis focuses on well-known ports for HTTP/HTTPS, email, and DNS services. We also include commonly used HTTPS alternative ports.
We use HTTPS scans from \muc, TLS scans for HTTPS alternative ports (including email and DNS ports) from Rapid7, and HTTP GET requests from Rapid7\footnote{Note that Rapid7 TLS scans are performed without \ac{sni} while our TLS scans from \muc include \acp{sni}.}.

\begin{table}
    \centering
    \footnotesize
    \caption{Application-layer responsiveness for detected \acp{hrp}. For TLS we use the certificate as identifier. For HTTP a hash of the response data.}
    \label{tbl:l7_result_stats}
    \begin{tabular}{lS[table-format=6.0]S[table-format=6.0]S[table-format=5.0]S[table-format=4.0]S[table-format=2.2]}
	    \toprule
	    & & \multicolumn{2}{c}{App. Layer Success} & \multicolumn{2}{c}{Same Identifier} \\
	    \cmidrule(lr){3-4}\cmidrule(lr){5-6}
        Port & {\# \acp{hrp}} & {\# \acp{hrp}} & {$>$\sperc{90} Success} & {\# \acp{hrp}} & {\ac{hrp} [\%]} \\
	    \midrule
	    80 & 91674 & 79234 & 64539 & 760 & 0.01 \\
	    443 &  64435 & 54203 & 26715 & 2718 & 0.1 \\
	    \midrule
	    1443 & 14461 & 557 & 364 & 33 & 9.1 \\
	    4443 & 9722 & 728 & 474 & 141 & 29.7 \\
	    8443 & 13048 & 3287 & 809 & 384 & 47.5 \\
	    \midrule
	    25 & 33294 & 3493 & 2210 & 2041 & 92.4 \\
	    110 & 11394 & 2553 & 2379 & 1944 & 81.7 \\
	    143 & 11112 & 2727 & 2527 & 2082 & 82.4 \\
	    465 & 9076 & 2793 & 2627 & 2174 & 82.8 \\
	    587 & 11274 & 2067 & 1056 & 931 & 88.2 \\
	    993 & 10362 & 2906 & 2703 & 2266 & 83.8 \\
	    995 & 10717 & 2910 & 2736 & 2303 & 84.2 \\
	    \midrule
	    853 & 8352 & 565 & 379 & 53 & 14.0 \\
	    \bottomrule
	\end{tabular}
\end{table}

\Cref{tbl:l7_result_stats} shows the results of application-layer scans for targets within our detected \acp{hrp}.
We differentiate between \first the availability of the expected application-layer service on at least one address and \second
the availability of the service on more than \sperc{90} of all previously responsive addresses.

For \sperc{86} of TCP/80 and \sperc{84} of TCP/443 \acp{hrp}, we find at least one host with a successful application-layer handshake.
\acp{cdn} play a major role as the top five organizations cover \sperc{64} of \acp{hrp}.
For \sk{27.7} of TCP/443 \acp{hrp}, or \sperc{49} of application-layer responsive \acp{hrp}, more than \sperc{90} of the respective addresses inside each prefix are responsive.
For the latter category, the fraction of \acp{hrp} with HTTP on TCP/80 is substantially higher (\sperc{81}) than that for TLS on TCP/443.
We note, however, that scanning a port with a \ac{tls} scanner can result in a specific \ac{tls} error, \eg when a \ac{sni} is not given (as is the case with the scans by Rapid7), even though
the corresponding hosts still offer the expected application-layer service in principle.
\reviewfix{B.9} When disregarding the targets with such errors\footnote{We omit these cases from \Cref{tbl:l7_result_stats} in favor of successful connections to provide useful references for the single identifier evaluation.}, the share of $>$ \sperc{90} application-layer service \acp{hrp} in TCP/443 is slightly higher (\sperc{89}) than that for TCP/80.

We evaluate the application-layer scans to determine if we find different answers from IP addresses inside an \ac{hrp}.
For TLS, we use the certificate as an indicator (we do not collect HTML responses at \muc).
Different certificates may indicate that these \acp{hrp} are unlikely to be aliased prefixes.
However, different answers are not necessarily indicators of the presence of distinct hosts, either \cite{alt2014v4aliases}.
To compare HTTP responses of different targets, we use a hash of the returned HTML responses.
This is more error-prone due to session-specific values such as the HTTP \texttt{Date} header or cookies.
Therefore, the results for HTTP must be interpreted as a lower bound.

\Cref{tbl:l7_result_stats} shows that only \sperc{0.01} of \acp{hrp} for port 80 and \sperc{0.1} of \acp{hrp} for port 443 use the same application-layer identifier within the prefix.
This is expected, as many \acp{cdn} offer HTTPS services, which are unlikely to return unique certificates, especially when scanned with \acs{sni} \cite{gasser2018log}.
In contrast, the TLS scans for email ports reveal that \sperc{80} of \acp{hrp} serve a single certificate on all scanned addresses.
\textit{home.pl}, a large polish web and mail hoster, is responsible for more than \sperc{30} of such \acp{hrp}.
We also find other similar, smaller providers within this group.

In comparison to the officially assigned HTTPS port 443, the alternative ports have a much smaller fraction of \acp{hrp} providing an application-layer service.
Only \sperc{4}--\sperc{25} of \acp{hrp} respond successfully on at least one address.
We also find a substantially higher proportion of identical certificates for alternative HTTPS ports in the ``$>$ \sperc{90} successful'' category.

\reviewfix{B.9}We also compare application-layer success of scanned targets within and outside \acp{hrp} in \Cref{tbl:l7_success_stats}.%
It clearly shows that our scans succeed more often for addresses not located within an \ac{hrp}.
In the case of TCP/80 and TCP/443 we find about \sperc{89} of addresses outside \acp{hrp} serving the expected service, while inside \acp{hrp} it is only \sperc{61} for TCP/443 and \sperc{75} for TCP/80.
For addresses inside \acp{hrp} we analyze the subset of prefixes where more than \sperc{90} of addresses expose the expected service.
This subset is responsible for more than half of all successful responses.
Digging further in this subset we find that ports related to mail services often seem to use the same certificate.
Therefore, a scan covering all responsive addresses not only often fails for \ac{hrp} addresses, but when a connection can actually be established the information obtained is almost entirely redundant.
This data also reflects our results from \Cref{ssec:as_analysis}, where we find two causes for \acp{hrp}: TCP proxies and \ac{cdn} deployments of Web services.

\begin{table}
    \centering
    \footnotesize
    \caption{Comparison of successful application-layer connections between non-\ac{hrp} and \ac{hrp} addresses. The last two columns represent the relative shares compared to the preceding column, with their subset (App. Layer $>$ \sperc{90}) containing addresses within \acp{hrp} meeting the \sperc{90} successful connections criterion. Note: This data is on addresses while \Cref{tbl:l7_result_stats} is on prefixes.}
    \label{tbl:l7_success_stats}
    \begin{tabular}{lS[table-format=2.1, table-space-text-post = \si{\percent}]S[table-format=2.1, table-space-text-post = \si{\percent}]S[table-format=2.1, table-space-text-post = \si{\percent}]S[table-format=2.1, table-space-text-post = \si{\percent}]}
        \toprule
        &&& \multicolumn{2}{c}{App. Layer $>$ \sperc{90}} \\
        \cmidrule(lr){4-5}
        Port & {Non-\ac{hrp}} & {\ac{hrp}} & {\ac{hrp} Share} & {Same Identifier} \\
        \midrule
        80  & 89.1 \si{\percent} & 73.5 \si{\percent} & 95.4 \si{\percent} & 1.2 \si{\percent} \\
        443 & 88.3 \si{\percent} & 61.1 \si{\percent} & 64.9 \si{\percent} & 10.6 \si{\percent} \\
        \midrule
        25  & 32.7 \si{\percent} &  6.7 \si{\percent} & 97.9 \si{\percent} & 92.3 \si{\percent} \\
        110 & 59.8 \si{\percent} & 20.8 \si{\percent} & 99.2 \si{\percent} & 81.9 \si{\percent} \\
        143 & 61.3 \si{\percent} & 22.6 \si{\percent} & 99.5 \si{\percent} & 82.6 \si{\percent} \\
        465 & 65.9 \si{\percent} & 28.7 \si{\percent} & 99.5 \si{\percent} & 82.9 \si{\percent} \\
        587 & 34.9 \si{\percent} & 12.7 \si{\percent} & 73.2 \si{\percent} & 87.4 \si{\percent} \\
        993 & 70.9 \si{\percent} & 26.1 \si{\percent} & 99.1 \si{\percent} & 83.7 \si{\percent} \\
        995 & 69.4 \si{\percent} & 25.4 \si{\percent} & 99.4 \si{\percent} & 84.1 \si{\percent} \\
        \midrule
        1443 & 7.0 \si{\percent} & 2.8 \si{\percent} & 88.4 \si{\percent} & 9.1 \si{\percent} \\
        4443 & 24.2 \si{\percent} & 5.3 \si{\percent} & 90.6 \si{\percent} & 29.7 \si{\percent} \\
        8443 & 67.7 \si{\percent} & 11.0 \si{\percent} & 55.2 \si{\percent} & 46.8 \si{\percent} \\
        \bottomrule
    \end{tabular}
\end{table}

\begin{figure}
    \includegraphics{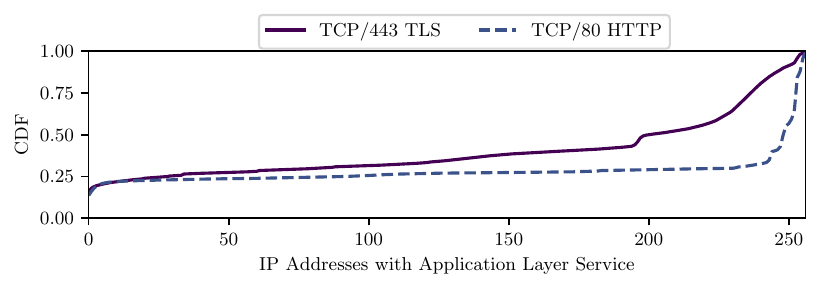}
    \caption{Cumulative distribution of IP addresses providing a TLS or HTTP service within \acp{hrp}.}
    \label{fig:l7_reachable_addrs_pfx}
\end{figure}

\Cref{fig:l7_reachable_addrs_pfx} shows the \ac{cdf} of \acp{hrp} with successful application-layer scans for ports 80 and 443.
Both protocols have close to \sperc{20} \acp{hrp} with unsuccessful application handshakes.
While TLS has \sperc{41} of \acp{hrp} with more than \sperc{90} availability, more than \sperc{82} of HTTP \acp{hrp} have at most \num{40} successful application-layer handshakes.
We also find \sperc{16} of application-layer responsive \acp{hrp} with more than \sperc{90} successful HTTP connection attempts.
Other TLS ports show a distribution similar to that for HTTP, with the difference that $>$ \sperc{80} of \acp{hrp} do not respond on a single address (see also \Cref{tbl:l7_result_stats}).

Finally, we also analyzed published data by \citeauthor{izhikevich2021lzr} \cite{izhikevich2021lzr}.
They provide the dataset where they scanned a sample of the IPv4 address space.
In total, the published dataset contains data on \sm{142} port scans for \sk{307} IP addresses.
We compare these scans with our \acp{hrp} on all available ports and locations from August 2022.
As our dataset contains only data on 142 ports we approximate if a prefix could be highly responsive on the scanned port by checking if it was classified as \ac{hrp} at a location for ten or more ports (see \Cref{ssec:port_eval}).
\sm{50} LZR scans are within this threshold and a comparison of scans within and outside this set reveals a significantly better response rate (\sperc{63} vs \sperc{91}) for scans outside \acp{hrp}.

\noindent
\textit{\textbf{Key take-away:}
When analyzing application-layer responsiveness, we observe a relatively higher share of responsive hosts outside \acp{hrp} as opposed to those within.
The application-layer analysis of deployment-specific properties reveals that alternative Web ports and mail ports often share a single identifier across an \ac{hrp}.
Finally, our validation on external scanning data reaffirms that application-layer scans performed on targets outside \acp{hrp} yield a higher success rate compared to targets within.
}

\subsection{DNS Analysis}
\label{ssec:dns}

We use DNS data provided by the OpenINTEL project to evaluate the theoretical population and addressing diversity of \acp{hrp}.
As we evaluate DNS data, we note that the information provided in DNS records may often come from a party other than the owner of a HRP; this is quite different from active scans into \acp{hrp}, which establish ground truth.
Theoretically, anyone can create DNS \texttt{A} records pointing to a specific address.
However, it is plausible to assume most DNS records express specific operational choices and may thus help to reveal the nature of an \ac{hrp}.
We use \ac{dns} results in our HTTPS scans and discuss its effect in \Cref{sec:discussion}.

\begin{table}
    \centering
    \footnotesize
    \caption{DNS statistics derived from \texttt{NS} records for \ac{dns} services and \texttt{MX} records for mail services with names which resolve to an IP address inside \acp{hrp}. The \acsp{sld} column represents the number of \acsp{sld} using such \texttt{NS}/\texttt{MX} records.}
    \label{tbl:oidnsstats}
    \begin{tabular}{lS[table-format=2.1, table-space-text-post = \si{\percent}]S[table-format=3.1, table-space-text-post = \si{\kn}]S[table-format=1.1, table-space-text-post = \si{\percent}]S[table-format=3.1, table-space-text-post = \si{\mn}]S[table-format=3.1, table-space-text-post = \si{\mn}]}
    \toprule
    {} & {\ac{hrp}} & \multicolumn{2}{c}{IP addresses}& {FQDNs} & {\acsp{sld}} \\
    \midrule
    TCP/53 & 12.0 \si{\percent}  & 40.9 \si{\kn} & 1.4 \si{\percent}  & 161.6 \si{\kn} & 115.6 \si{\mn}  \\
    UDP/53 & 25.5 \si{\percent}  & 29.0 \si{\kn} & 3.1 \si{\percent}  & 133.0 \si{\kn} & 104.6 \si{\mn} \\
    TCP/853 & 2.3 \si{\percent}  & 1.0 \si{\kn} & 0.1 \si{\percent}  & 1.2 \si{\kn} & 55.2 \si{\kn} \\
    \midrule
    \multicolumn{5}{l}{\textit{Mail Ports:}} \\
    \midrule
    TCP/25  & 18.5 \si{\percent} & 172.4 \si{\kn} & 2.0 \si{\percent} & 3.0 \si{\mn} & 3.7 \si{\mn} \\
    TCP/110 & 26.4 \si{\percent} & 126.0 \si{\kn} & 4.4 \si{\percent} & 2.7 \si{\mn} & 3.2 \si{\mn} \\
    TCP/143 & 26.3 \si{\percent} & 121.6 \si{\kn} & 4.3 \si{\percent} & 2.7 \si{\mn} & 3.2 \si{\mn} \\
    TCP/465 & 30.4 \si{\percent} & 121.0 \si{\kn} & 5.3 \si{\percent} & 2.7 \si{\mn} & 3.2 \si{\mn} \\
    TCP/587 & 30.6 \si{\percent} & 133.1 \si{\kn} & 4.7 \si{\percent} & 2.7 \si{\mn} & 3.2 \si{\mn} \\
    TCP/993 & 27.2 \si{\percent} & 120.0 \si{\kn} & 4.6 \si{\percent} & 2.7 \si{\mn} & 3.2 \si{\mn} \\
    TCP/995 & 27.5 \si{\percent} & 261.5 \si{\kn} & 4.6 \si{\percent} & 5.6 \si{\mn} & 6.6 \si{\mn} \\
    \bottomrule
\end{tabular}
\end{table}

\Cref{tbl:oidnsstats} shows how many fully qualified domain names (FQDNs) and \acp{sld} are associated with specific ports and protocols (DNS and email) on average.
We consider the respective record types (\ie \texttt{NS} for DNS and
\texttt{MX} for email) and map a name server or mail server to \acp{hrp}
if at least one of its \texttt{A} records points to an address within the
\acp{hrp}.
For mail services we assume a host used for \texttt{MX} records also provides POP3 or IMAP4 services.
The table also shows the number of unique IP addresses contained in \texttt{A}
records, and how many domain names reference the associated name servers and mail
exchangers.

UDP/53 \acp{hrp} are usually referred to by DNS \texttt{NS} records.
While only about \sperc{3} of possible \ac{hrp} addresses are actually present
in name server resolutions, these name servers serve more than \sm{100}
\acp{sld}.
The smaller results for TCP/853 are expected as DNS-over-TLS is not very widely deployed
and is more commonly used by resolvers, not nameservers.
If we consider the IP address space covered by DNS \acp{hrp}, we observe that
some \ac{hrp} are referenced on at least 90\% of their addresses by some
\texttt{NS} record. Specifically, this involves 28 TCP/53 and 15 UDP/53
\acp{hrp}.
Although an open port is
not necessarily indicative of service availability, DNS references to
\ac{hrp} addresses as well as open DNS ports therein suggest that DNS services are present in
the respective \ac{hrp}.

We find more \texttt{MX} records that point into \acp{hrp} than \texttt{NS} records.
There are also significantly more unique \texttt{MX}
FQDNs than \texttt{NS} FQDNs. However, the number of \acsp{sld} related to \texttt{MX} records is only about the same number as the distinct \texttt{MX} record names.
Depending on the port, 1\% to 2\% of \acp{hrp} have at least 90\% of their respective addresses referenced via \texttt{MX} records.

\begin{table}
    \centering
    \footnotesize
    \caption{Unique number of IP addresses, the fraction of \ac{hrp} address space they cover, and FQDNs derived from \texttt{A} records for HTTP/HTTPS ports. }
    \label{tbl:http_dns_stats}
    \begin{tabular}{lS[table-format=2.1, table-space-text-post = \si{\percent}]S[table-format=3.1, table-space-text-post = \si{\kn}]S[table-format=2.1, table-space-text-post = \si{\percent}]S[table-format=3.1, table-space-text-post = \si{\mn}]}
    \toprule
    {} & {\ac{hrp}}  &  \multicolumn{2}{c}{IP addresses} & {FQDNs} \\
    \midrule
    TCP/80   & 34.4 \si{\percent} &   4.7 \si{\mn} & 11.0 \si{\percent} & 171.4 \si{\mn} \\
    TCP/443  & 30.8 \si{\percent} &   2.0 \si{\mn} & 6.3 \si{\percent} & 149.1 \si{\mn} \\
    TCP/8000 & 32.7 \si{\percent} & 349.2 \si{\kn} & 10.9 \si{\percent} & 2.1 \si{\mn} \\
    TCP/8080 & 45.9 \si{\percent} & 504.2 \si{\kn} & 12.6 \si{\percent} & 27.3 \si{\mn} \\
    TCP/8443 & 56.8 \si{\percent} & 517.3 \si{\kn} & 16.7 \si{\percent} & 28.1 \si{\mn} \\
    \bottomrule
\end{tabular}

\end{table}

We derive possible use of HTTP/HTTPS from DNS data as well by inspecting
the \texttt{A} records for domain name's apexes. \Cref{tbl:http_dns_stats}
summarizes the results. Of the port-specific HTTP \acp{hrp}, \sperc{30}---\sperc{57} are
related to an \texttt{A} record in the DNS. Nearly half of all \acp{hrp} with
popular alternative ports are referenced in the DNS data.  Interestingly, fewer specific IP addresses
in TCP/443 \acp{hrp} are referenced compared to TCP/80 \acp{hrp}.

A large \ac{cdn} presence is also visible in the \ac{dns} data.
\sm{149} FQDNs resolve to TCP/443 \acp{hrp} and \sm{171} to TCP/80 \acp{hrp}.
With this in mind, it is not surprising to find Alexa Top 1M
domains resolving to \sk{366} IP addresses inside TCP/443 \acp{hrp} and \sk{377} IP addresses inside TCP/80 \acp{hrp}.
Therefore, \ac{dns} data shows that while only few IP addresses are referenced, these are used by many millions of FQDNs.
Nevertheless, more than \sperc{80} of the address space is not referred to by \ac{dns}.
Hence, domain-based measurement studies hit only a fraction of addresses compared to seeding the input from port scans.

\noindent
\textit{\textbf{Key take-away:}
	The analysis of \ac{dns} data provided by OpenINTEL shows that only a fraction of responsive addresses is actually referred to in \ac{dns}.
    Millions of FQDNs resolve to addresses inside \acp{hrp} or refer to \texttt{MX} and \texttt{NS} domains resolving to addresses inside \acp{hrp}.
    Therefore, \acp{hrp} can contain important hosts, but the number of referred addresses is smaller and can be identified via \ac{dns} resolutions.
}
\section{HRP-Aware Scanning Technique}
\label{sec:discussion}

We evaluated the presence of \acp{hrp} and showed that \ac{hrp} addresses can
account for up to \sperc{80} of all responsive addresses, especially in the
case of non-well-known ports.
We provided evidence that \ac{hrp} addresses often expose few or no application-layer
services at all.  Our findings lead us to conclude that many prefixes are
of limited value for scans and can incur bias in the results.
At the same time, our work confirms---and gives broader and firmer
evidence---for a previous result, namely that simple port scans are
insufficient to determine service availability and liveness.  The conclusion is
that while application-layer scans are necessary for the reliable assessment of
service deployments, the approach to carrying these out needs to be optimized.

Internet-wide application-layer scans typically commence with a port scan (e.g., employing ZMap) to identify responsive addresses.
Subsequently, an application-layer scan is executed, focusing solely on the IP addresses that have responded to the port scan.
Application-layer scans are more time-consuming due to extended message exchanges and additional round-trips, and they also impose a heavier load on the scanned entities.
Therefore, it becomes imperative to minimize the scanning scope to what is genuinely essential.
One way to achieve this optimization is to consider \acp{hrp} before initiating the scan.
Consequently, our proposal aims to align with this objective. \reviewfix{F.1}

\subsection{Approach}
We suggest using a list of \acp{hrp}, as produced by our
tool, and scan these prefixes strategically.
The use case will determine the exact strategy, but the following methods can be used and combined.
If available, researchers should use data from DNS resolutions to determine IP
addresses where services might be deployed inside an \ac{hrp}.
Scanning based on domain names is important in a number of cases, but
especially for \ac{tls} scans, where omitting the \ac{sni} may lead to
inaccurate results.
If a use case cannot be addressed by scanning based on
resolved domain names, we suggest a hybrid approach: instead of scanning
\acp{hrp} in full, we suggest taking a sample that consists of IP
addresses that are associated to DNS records (\texttt{A}, \texttt{NS}, \texttt{MX},
etc.) plus addresses sampled uniformly from the remainder of the \ac{hrp}.
The selection of appropriate domain names, by using the \texttt{SVCB/HTTPS} DNS records~\cite{zirngibl2023svcb} to find service names and excluding parked domains~\cite{zirngibl2022prevalenceofparking}, can improve the obtained results.
If, for example, we decide to take a sample of ten addresses per \ac{hrp}, and
we find only five different IP addresses in \ac{dns}, we suggest filling up the
remaining five addresses through sampling.

The sampled data will yield results that can be used to make further decisions.
We identify three scenarios: \first the \ac{hrp} seems to be a proxy, \second
the \ac{hrp} behaves like a \ac{cdn} and always returns the same result (\eg
the same \ac{tls} certificate), and \third different application-layer
responses for different hosts are returned.
In the first two scenarios, the research question must be considered
to decide whether further samples are necessary for statistically strong
results, or whether the sample data is already enough.  The third scenario
describes an \ac{hrp} which we suggest scanning in full.

\subsection{Evaluation}
We evaluate the effect of this approach on our
own \ac{tls} scans by differentiating between the \ac{sni} scan which uses
\ac{dns} resolutions as input list and the non-\ac{sni} scan which uses plain
IP addresses from the port scan.
We find that \sperc{30} of the \ac{sni} scan targets, which can be successfully scanned, are within an \ac{hrp}, while \sperc{80} of targets in the non-\ac{sni} scan are. \reviewfix{B.9}
In \Cref{ssec:dns} we showed that \ac{dns} scans provide a high quality input
list, but we find that these also omit responsive \acp{hrp}.  Therefore,
applying a sampling rate on \acp{hrp} which are not covered by \ac{dns} is
highly beneficial.  The sampling step also helps to prevent unnecessary
application-layer handshakes in case of redundant responses.  A trivial
sampling algorithm which selects ten addresses uniformly from a \texttt{/24}
successfully completes connections to \sperc{92} of \acp{hrp} compared to the
full scan.

A similar evaluation of Rapid7 TLS data for TLS alternative ports and other TLS-enables services, serves to validate our observations.
Rapid7's dataset is not as detailed as our own scans as they only provide us data of successful scans, and exclusively perform TLS scans without \ac{sni}.
However, by approximating the number of scans based on their TCP port scans, typically conducted a few days prior to the TLS scans, we can derive the missing numbers considering they use the standard approach (ZMap followed by the application layer scan).
Our proposed approach yields even more favorable results for these additional ports, as it still collects over \sperc{99} of unique certificates while reducing application layer scans by up to \sperc{75} depending on the specific ports being assessed (number based on evaluations in \Cref{ssec:port_eval} and \Cref{ssec:app_layer_results}). \reviewfix{D.7}
These findings demonstrate the broad applicability and efficacy of our approach, not being limited to specific ports.

\subsection{Approach in Practice}
Our \ac{hrp} filter tool needs the full port
scan results to determine which prefixes are highly responsive.  Therefore, one
option is to run the port scan first, then apply our filter, and only
afterwards starting the application-layer scan.  One can also use a previous
scan result or the weekly \ac{hrp} results which we publish. Either will work
as we have shown the stability of \ac{hrp} in time (see \Cref{ssec:stability})
and space (see \Cref{ssec:bias}).
Therefore, the approach also applicable to data sources such as Rapid7's Project Sonar~\cite{rapid7} and Censys~\cite{durumeric_censys_2015} port scans.
We note that \acp{hrp} can be filtered during scanning and hence would not interfere when the timing of application-layer scans is critical.

\subsection{Pros and Cons}
Our approach offers two substantial advantages.
It has the potential to reduce scanning effort by \sperc{20}-\sperc{80}, as determined through our \ac{hrp} analysis.
Especially many of the \ac{tls} alternative port \ac{hrp} addresses exhibit a low success rate.
Given that timeouts are typically the most time-consuming aspect during \ac{tls} handshakes, their exclusion provides a substantial benefit.
Nevertheless, our approach is not without limitations.
First and foremost, it cannot ensure the comprehensive scanning of all reachable, distinct systems.
This limitation may pose challenges when the scanning objective is to identify every individual host with a particular configuration, as is often the case in certain vulnerability scans.
However, it is essential to recognize that when the research goal centers around statistical aggregates, our approach contributes to reducing bias and is advantageous for the overall research outcome.
Based on our results, we believe that many application-layer scanning campaigns
can lower their impact on the network and improve their results by considering
our approach.

\section{Ethical Considerations}
\label{sec:ethics}

We apply strict ethical measures to all our scans~\cite{menloreport} and adopt
community best practices~\cite{PA16}.
\reviewfix{B.3}
We uphold this ethical approach by imposing limitations on our scan rate and maintaining an internal blocklist shared by all vantage points to record opt-out requests.
Consequently, if a party requests not to be scanned, all vantage points will cease scanning their systems.
Furthermore, each scanning machine identifies itself through reverse DNS, a hosted website, and informative \texttt{WHOIS} entries.
Besides conducting all scans ourselves, we verify that findings are similar from different vantage points (see \Cref{ssec:bias}).
\reviewfix{B.3}
This enables us to leverage pre-existing data from Rapid7 for the majority of port and application-layer scans.
Furthermore, scans conducted across multiple vantage points are less likely to trigger intrusion detection systems, thanks to the reduced load from a single scanning entity.
Consequently, they are less disruptive for network operators.
\reviewfix{B.3}
Similar measures are applied by Rapid7: They publish their scanning prefixes, have an opt-out process, and describe more relevant considerations on their website~\cite{rapid7considerations}.
\section{Conclusion}
\label{sec:conclusion}

In this paper we developed and presented a technique to detect \acf{hrp} and
analyzed the presence of \acp{hrp} on 142 TCP and 19 UDP ports in the IPv4 Internet.
We developed a tool to detect \acp{hrp} and used it to analyze port scans for 142 TCP ports.
We found that \acp{hrp} are highly visible on any analyzed TCP port and can make up to \sperc{80} of responsive addresse.
\acp{cdn} are the largest entities exposing \acp{hrp} for well-known Web ports.
In some cases, \acp{cdn} deploy reverse proxies, which results in
\acp{hrp} responsivity on all available ports.
We proposed different approaches towards more efficient scanning techniques, depending on the research goals in question and showed that these can significantly reduce the number of targets to be scanned.
Finally, we released tooling to this end, and will publish data and statistics on \acp{hrp} going forward.

\noindent\textbf{Acknowledgements.} We thank the anonymous reviewers and our shepherd for their valuable feedback. This work was partially funded by the German Federal Ministry of Education and Research under project PRIMEnet (16KIS1370) and the Netherlands Organisation for Scientific Research project CATRIN (NWA.1215.18.003).
Additionally, we received funding by the European Union's Horizon 2020 research and innovation program (grant agreement no. 101008468 and 101079774) as well as the German Research Foundation (HyperNIC, grant no. CA595/13-1).
This research was made possible by OpenINTEL, a joint project of the University of Twente, SURF, SIDN, and NLnet Labs.

\bibliographystyle{ACM-Reference-Format}
\bibliography{paper}


\begin{thebibliography}{43}


\ifx \showCODEN    \undefined \def \showCODEN     #1{\unskip}     \fi
\ifx \showDOI      \undefined \def \showDOI       #1{#1}\fi
\ifx \showISBNx    \undefined \def \showISBNx     #1{\unskip}     \fi
\ifx \showISBNxiii \undefined \def \showISBNxiii  #1{\unskip}     \fi
\ifx \showISSN     \undefined \def \showISSN      #1{\unskip}     \fi
\ifx \showLCCN     \undefined \def \showLCCN      #1{\unskip}     \fi
\ifx \shownote     \undefined \def \shownote      #1{#1}          \fi
\ifx \showarticletitle \undefined \def \showarticletitle #1{#1}   \fi
\ifx \showURL      \undefined \def \showURL       {\relax}        \fi
\providecommand\bibfield[2]{#2}
\providecommand\bibinfo[2]{#2}
\providecommand\natexlab[1]{#1}
\providecommand\showeprint[2][]{arXiv:#2}

\bibitem[Alt et~al\mbox{.}(2014)]%
        {alt2014v4aliases}
\bibfield{author}{\bibinfo{person}{Lance Alt}, \bibinfo{person}{Robert
  Beverly}, {and} \bibinfo{person}{Alberto Dainotti}.}
  \bibinfo{year}{2014}\natexlab{}.
\newblock \showarticletitle{{Uncovering Network Tarpits with Degreaser}}. In
  \bibinfo{booktitle}{\emph{Proceedings of the 30th Annual Computer Security
  Applications Conference}} (New Orleans, Louisiana, USA).
\newblock


\bibitem[Bano et~al\mbox{.}(2018)]%
        {bano2018liveness}
\bibfield{author}{\bibinfo{person}{Shehar Bano}, \bibinfo{person}{Philipp
  Richter}, \bibinfo{person}{Mobin Javed}, \bibinfo{person}{Srikanth
  Sundaresan}, \bibinfo{person}{Zakir Durumeric}, \bibinfo{person}{Steven~J.
  Murdoch}, \bibinfo{person}{Richard Mortier}, {and} \bibinfo{person}{Vern
  Paxson}.} \bibinfo{year}{2018}\natexlab{}.
\newblock \showarticletitle{{Scanning the Internet for Liveness}}.
\newblock \bibinfo{journal}{\emph{ACM SIGCOMM Computer Communication Review}}
  (\bibinfo{year}{2018}).
\newblock


\bibitem[Beverly et~al\mbox{.}(2013)]%
        {beverly2013tbt}
\bibfield{author}{\bibinfo{person}{Robert Beverly}, \bibinfo{person}{William
  Brinkmeyer}, \bibinfo{person}{Matthew Luckie}, {and}
  \bibinfo{person}{Justin~P. Rohrer}.} \bibinfo{year}{2013}\natexlab{}.
\newblock \showarticletitle{{IPv6 Alias Resolution via Induced Fragmentation}}.
  In \bibinfo{booktitle}{\emph{Proc. Passive and Active Measurement (PAM)}}.
\newblock


\bibitem[Cloudflare(2019)]%
        {cfspectrumblog}
\bibfield{author}{\bibinfo{person}{Cloudflare}.}
  \bibinfo{year}{2019}\natexlab{}.
\newblock \bibinfo{booktitle}{\emph{{It's crowded in here!}}}
\newblock
\urldef\tempurl%
\url{https://blog.cloudflare.com/its-crowded-in-here/}
\showURL{%
\tempurl}


\bibitem[Cloudflare(2021)]%
        {cfaddressagility}
\bibfield{author}{\bibinfo{person}{Cloudflare}.}
  \bibinfo{year}{2021}\natexlab{}.
\newblock \bibinfo{booktitle}{\emph{{Unbuckling the narrow waist of IP:
  Addressing Agility for Names and Web Services}}}.
\newblock
\urldef\tempurl%
\url{https://blog.cloudflare.com/addressing-agility/}
\showURL{%
\tempurl}


\bibitem[Cloudflare(2023a)]%
        {cfspectrum}
\bibfield{author}{\bibinfo{person}{Cloudflare}.}
  \bibinfo{year}{2023}\natexlab{a}.
\newblock \bibinfo{booktitle}{\emph{{Cloudflare Spectrum}}}.
\newblock
\urldef\tempurl%
\url{https://www.cloudflare.com/products/cloudflare-spectrum/}
\showURL{%
\tempurl}


\bibitem[Cloudflare(2023b)]%
        {cfspectrumports}
\bibfield{author}{\bibinfo{person}{Cloudflare}.}
  \bibinfo{year}{2023}\natexlab{b}.
\newblock \bibinfo{booktitle}{\emph{{Cloudflare Spectrum - Network ports}}}.
\newblock
\urldef\tempurl%
\url{https://developers.cloudflare.com/fundamentals/get-started/reference/network-ports/}
\showURL{%
\tempurl}


\bibitem[Costin et~al\mbox{.}(2014)]%
        {costin2014embedded}
\bibfield{author}{\bibinfo{person}{Andrei Costin}, \bibinfo{person}{Jonas
  Zaddach}, \bibinfo{person}{Aur{\'e}lien Francillon}, {and}
  \bibinfo{person}{Davide Balzarotti}.} \bibinfo{year}{2014}\natexlab{}.
\newblock \showarticletitle{{A Large-Scale Analysis of the Security of Embedded
  Firmwares}}. In \bibinfo{booktitle}{\emph{23rd USENIX Security Symposium
  (USENIX Security 14)}}. \bibinfo{publisher}{USENIX Association},
  \bibinfo{address}{San Diego, CA}, \bibinfo{pages}{95--110}.
\newblock


\bibitem[Dittrich et~al\mbox{.}(2012)]%
        {menloreport}
\bibfield{author}{\bibinfo{person}{David Dittrich}, \bibinfo{person}{Erin
  Kenneally}, {et~al\mbox{.}}} \bibinfo{year}{2012}\natexlab{}.
\newblock \showarticletitle{{The Menlo Report: Ethical principles guiding
  information and communication technology research}}.
\newblock \bibinfo{journal}{\emph{US Department of Homeland Security}}
  (\bibinfo{year}{2012}).
\newblock


\bibitem[Durand et~al\mbox{.}(2015)]%
        {rfc7454}
\bibfield{author}{\bibinfo{person}{J. Durand}, \bibinfo{person}{I. Pepelnjak},
  {and} \bibinfo{person}{G. Doering}.} \bibinfo{year}{2015}\natexlab{}.
\newblock \bibinfo{title}{{BGP Operations and Security}}.
\newblock \bibinfo{howpublished}{RFC 7454 (Best Current Practice)}.
\newblock
\showISSN{2070-1721}
\urldef\tempurl%
\url{https://doi.org/10.17487/RFC7454}
\showDOI{\tempurl}


\bibitem[Durumeric et~al\mbox{.}(2015)]%
        {durumeric_censys_2015}
\bibfield{author}{\bibinfo{person}{Zakir Durumeric}, \bibinfo{person}{David
  Adrian}, \bibinfo{person}{Ariana Mirian}, \bibinfo{person}{Michael Bailey},
  {and} \bibinfo{person}{J.~Alex Halderman}.} \bibinfo{year}{2015}\natexlab{}.
\newblock \showarticletitle{{A Search Engine Backed by Internet-Wide
  Scanning}}. In \bibinfo{booktitle}{\emph{Proceedings of the 22nd ACM SIGSAC
  Conference on Computer and Communications Security}} (Denver, Colorado, USA)
  \emph{(\bibinfo{series}{CCS '15})}. \bibinfo{publisher}{Association for
  Computing Machinery}, \bibinfo{address}{New York, NY, USA},
  \bibinfo{pages}{542–553}.
\newblock


\bibitem[Durumeric et~al\mbox{.}(2014)]%
        {durumeric_heartbleed_2014}
\bibfield{author}{\bibinfo{person}{Zakir Durumeric}, \bibinfo{person}{Frank
  Li}, \bibinfo{person}{James Kasten}, \bibinfo{person}{Johanna Amann},
  \bibinfo{person}{Jethro Beekman}, \bibinfo{person}{Mathias Payer},
  \bibinfo{person}{Nicolas Weaver}, \bibinfo{person}{David Adrian},
  \bibinfo{person}{Vern Paxson}, \bibinfo{person}{Michael Bailey}, {and}
  \bibinfo{person}{J.~Alex Halderman}.} \bibinfo{year}{2014}\natexlab{}.
\newblock \showarticletitle{{The Matter of Heartbleed}}. In
  \bibinfo{booktitle}{\emph{Proceedings of the 2014 Conference on Internet
  Measurement Conference}} (Vancouver, BC, Canada) \emph{(\bibinfo{series}{IMC
  '14})}. \bibinfo{publisher}{Association for Computing Machinery},
  \bibinfo{address}{New York, NY, USA}, \bibinfo{pages}{475–488}.
\newblock


\bibitem[Durumeric et~al\mbox{.}(2013)]%
        {durumeric_zmap_2013}
\bibfield{author}{\bibinfo{person}{Zakir Durumeric}, \bibinfo{person}{Eric
  Wustrow}, {and} \bibinfo{person}{J.~Alex Halderman}.}
  \bibinfo{year}{2013}\natexlab{}.
\newblock \showarticletitle{{ZMap: Fast Internet-wide Scanning and Its Security
  Applications}}. In \bibinfo{booktitle}{\emph{Proc. USENIX Security
  Symposium}}. \bibinfo{address}{Washington, D.C., USA}.
\newblock


\bibitem[Fayed et~al\mbox{.}(2021)]%
        {fayed2021addressagility}
\bibfield{author}{\bibinfo{person}{Marwan Fayed}, \bibinfo{person}{Lorenz
  Bauer}, \bibinfo{person}{Vasileios Giotsas}, \bibinfo{person}{Sami Kerola},
  \bibinfo{person}{Marek Majkowski}, \bibinfo{person}{Pavel Odintsov},
  \bibinfo{person}{Jakub Sitnicki}, \bibinfo{person}{Taejoong Chung},
  \bibinfo{person}{Dave Levin}, \bibinfo{person}{Alan Mislove},
  \bibinfo{person}{Christopher~A. Wood}, {and} \bibinfo{person}{Nick
  Sullivan}.} \bibinfo{year}{2021}\natexlab{}.
\newblock \showarticletitle{{The Ties That Un-Bind: Decoupling IP from Web
  Services and Sockets for Robust Addressing Agility at CDN-Scale}}. In
  \bibinfo{booktitle}{\emph{Proceedings of the 2021 ACM SIGCOMM 2021
  Conference}} \emph{(\bibinfo{series}{SIGCOMM '21})}.
  \bibinfo{publisher}{Association for Computing Machinery},
  \bibinfo{address}{New York, NY, USA}.
\newblock


\bibitem[Gasser et~al\mbox{.}(2018a)]%
        {gasser2018log}
\bibfield{author}{\bibinfo{person}{Oliver Gasser}, \bibinfo{person}{Benjamin
  Hof}, \bibinfo{person}{Max Helm}, \bibinfo{person}{Maciej Korczynski},
  \bibinfo{person}{Ralph Holz}, {and} \bibinfo{person}{Georg Carle}.}
  \bibinfo{year}{2018}\natexlab{a}.
\newblock \showarticletitle{{In Log We Trust: Revealing Poor Security Practices
  with Certificate Transparency Logs and Internet Measurements}}. In
  \bibinfo{booktitle}{\emph{Passive and Active Measurement Conference 2018}}.
\newblock


\bibitem[Gasser et~al\mbox{.}(2018b)]%
        {gasser2018clusters}
\bibfield{author}{\bibinfo{person}{Oliver Gasser}, \bibinfo{person}{Quirin
  Scheitle}, \bibinfo{person}{Pawel Foremski}, \bibinfo{person}{Qasim Lone},
  \bibinfo{person}{Maciej Korczynski}, \bibinfo{person}{Stephen~D. Strowes},
  \bibinfo{person}{Luuk Hendriks}, {and} \bibinfo{person}{Georg Carle}.}
  \bibinfo{year}{2018}\natexlab{b}.
\newblock \showarticletitle{{Clusters in the Expanse: Understanding and
  Unbiasing IPv6 Hitlists}}. In \bibinfo{booktitle}{\emph{Proc. ACM Int.
  Measurement Conference (IMC)}} (Boston, MA, USA).
\newblock


\bibitem[Gasser et~al\mbox{.}(2016)]%
        {gasser2016scanning}
\bibfield{author}{\bibinfo{person}{Oliver Gasser}, \bibinfo{person}{Quirin
  Scheitle}, \bibinfo{person}{Sebastian Gebhard}, {and} \bibinfo{person}{Georg
  Carle}.} \bibinfo{year}{2016}\natexlab{}.
\newblock \showarticletitle{{Scanning the IPv6 Internet: Towards a
  Comprehensive Hitlist}}. In \bibinfo{booktitle}{\emph{Proc. 8th Int. Workshop
  on Traffic Monitoring and Analysis}}. \bibinfo{address}{Louvain-la-Neuve,
  Belgium}.
\newblock


\bibitem[Gasser et~al\mbox{.}(2023)]%
        {goscanner}
\bibfield{author}{\bibinfo{person}{Oliver Gasser}, \bibinfo{person}{Markus
  Sosnowski}, \bibinfo{person}{Patrick Sattler}, {and}
  \bibinfo{person}{Johannes Zirngibl}.} \bibinfo{year}{2023}\natexlab{}.
\newblock \bibinfo{booktitle}{\emph{{Goscanner}}}.
\newblock
\urldef\tempurl%
\url{https://github.com/tumi8/goscanner}
\showURL{%
Retrieved 2023-03-24 from \tempurl}


\bibitem[Graham({[n.\,d.]})]%
        {masscan}
\bibfield{author}{\bibinfo{person}{Robert Graham}.}
  \bibinfo{year}{[n.\,d.]}\natexlab{}.
\newblock \bibinfo{booktitle}{\emph{{MASSCAN: Mass IP port scanner}}}.
\newblock
\urldef\tempurl%
\url{https://github.com/robertdavidgraham/masscan}
\showURL{%
\tempurl}


\bibitem[Hofmann(2013)]%
        {rapid7considerations}
\bibfield{author}{\bibinfo{person}{Marcia Hofmann}.}
  \bibinfo{year}{2013}\natexlab{}.
\newblock \bibinfo{booktitle}{\emph{Legal Considerations for Widespread
  Scanning}}.
\newblock
\urldef\tempurl%
\url{https://www.rapid7.com/blog/post/2013/10/30/legal-considerations-for-widespread-scanning/}
\showURL{%
Retrieved 2023-09-26 from \tempurl}


\bibitem[https://csirt.divd.nl/(2023)]%
        {divd}
\bibfield{author}{\bibinfo{person}{https://csirt.divd.nl/}.}
  \bibinfo{year}{2023}\natexlab{}.
\newblock \bibinfo{booktitle}{\emph{Making the internet safer through
  Coordinated Vulnerability Disclosure}}.
\newblock
\urldef\tempurl%
\url{https://csirt.divd.nl/}
\showURL{%
Retrieved 2023-03-24 from \tempurl}


\bibitem[ICANN(2023)]%
        {icannczds}
\bibfield{author}{\bibinfo{person}{ICANN}.} \bibinfo{year}{2023}\natexlab{}.
\newblock \bibinfo{booktitle}{\emph{{CZDS - Centralized Zone Data Service}}}.
\newblock
\urldef\tempurl%
\url{https://czds.icann.org/}
\showURL{%
Retrieved 2023-10-05 from \tempurl}


\bibitem[Izhikevich et~al\mbox{.}(2021)]%
        {izhikevich2021lzr}
\bibfield{author}{\bibinfo{person}{Liz Izhikevich}, \bibinfo{person}{Renata
  Teixeira}, {and} \bibinfo{person}{Zakir Durumeric}.}
  \bibinfo{year}{2021}\natexlab{}.
\newblock \showarticletitle{{LZR}: Identifying Unexpected Internet Services}.
  In \bibinfo{booktitle}{\emph{Proc. USENIX Security Symposium}}.
\newblock
\urldef\tempurl%
\url{https://www.usenix.org/conference/usenixsecurity21/presentation/izhikevich}
\showURL{%
\tempurl}


\bibitem[Izhikevich et~al\mbox{.}(2022)]%
        {izhikevich2022predicting}
\bibfield{author}{\bibinfo{person}{Liz Izhikevich}, \bibinfo{person}{Renata
  Teixeira}, {and} \bibinfo{person}{Zakir Durumeric}.}
  \bibinfo{year}{2022}\natexlab{}.
\newblock \showarticletitle{Predicting IPv4 Services across All Ports}. In
  \bibinfo{booktitle}{\emph{Proceedings of the ACM SIGCOMM 2022 Conference}}
  (Amsterdam, Netherlands) \emph{(\bibinfo{series}{SIGCOMM '22})}.
  \bibinfo{publisher}{Association for Computing Machinery},
  \bibinfo{address}{New York, NY, USA}, \bibinfo{pages}{503–515}.
\newblock
\showISBNx{9781450394208}
\urldef\tempurl%
\url{https://doi.org/10.1145/3544216.3544249}
\showDOI{\tempurl}


\bibitem[Klick et~al\mbox{.}(2016)]%
        {klick2016tass}
\bibfield{author}{\bibinfo{person}{Johannes Klick}, \bibinfo{person}{Stephan
  Lau}, \bibinfo{person}{Matthias W\"{a}hlisch}, {and} \bibinfo{person}{Volker
  Roth}.} \bibinfo{year}{2016}\natexlab{}.
\newblock \showarticletitle{{Towards Better Internet Citizenship: Reducing the
  Footprint of Internet-Wide Scans by Topology Aware Prefix Selection}}. In
  \bibinfo{booktitle}{\emph{Proc. ACM Int. Measurement Conference (IMC)}}
  (Santa Monica, California, USA). \bibinfo{publisher}{Association for
  Computing Machinery}, \bibinfo{address}{New York, NY, USA}.
\newblock


\bibitem[Luckie et~al\mbox{.}(2013)]%
        {luckie2013speedtrap}
\bibfield{author}{\bibinfo{person}{Matthew Luckie}, \bibinfo{person}{Robert
  Beverly}, \bibinfo{person}{William Brinkmeyer}, {and} \bibinfo{person}{kc
  claffy}.} \bibinfo{year}{2013}\natexlab{}.
\newblock \showarticletitle{{Speedtrap: Internet-Scale IPv6 Alias Resolution}}.
  In \bibinfo{booktitle}{\emph{Proc. ACM Int. Measurement Conference (IMC)}}
  (Barcelona, Spain).
\newblock


\bibitem[MANRS(2021)]%
        {MANRS2021Filter}
\bibfield{author}{\bibinfo{person}{MANRS}.} \bibinfo{year}{2021}\natexlab{}.
\newblock \bibinfo{title}{Prefix filter configuration tools}.
\newblock
\newblock
\urldef\tempurl%
\url{https://www.manrs.org/isps/guide/filtering/}
\showURL{%
\tempurl}


\bibitem[Murdock et~al\mbox{.}(2017)]%
        {murdock20176Gen}
\bibfield{author}{\bibinfo{person}{Austin Murdock}, \bibinfo{person}{Frank Li},
  \bibinfo{person}{Paul Bramsen}, \bibinfo{person}{Zakir Durumeric}, {and}
  \bibinfo{person}{Vern Paxson}.} \bibinfo{year}{2017}\natexlab{}.
\newblock \showarticletitle{{Target Generation for Internet-Wide IPv6
  Scanning}}. In \bibinfo{booktitle}{\emph{Proc. ACM Int. Measurement
  Conference (IMC)}} (London, United Kingdom).
\newblock


\bibitem[of~Oregon(2023)]%
        {routeviews}
\bibfield{author}{\bibinfo{person}{University of Oregon}.}
  \bibinfo{year}{2023}\natexlab{}.
\newblock \bibinfo{booktitle}{\emph{{University of Oregon Route Views
  Project}}}.
\newblock
\urldef\tempurl%
\url{http://www.routeviews.org/routeviews/}
\showURL{%
\tempurl}


\bibitem[Padmanabhan et~al\mbox{.}(2015)]%
        {padmanabhan2015uav6}
\bibfield{author}{\bibinfo{person}{Ramakrishna Padmanabhan},
  \bibinfo{person}{Zhihao Li}, \bibinfo{person}{Dave Levin}, {and}
  \bibinfo{person}{Neil Spring}.} \bibinfo{year}{2015}\natexlab{}.
\newblock \showarticletitle{{UAv6: Alias Resolution in IPv6 Using Unused
  Addresses}}. In \bibinfo{booktitle}{\emph{Proc. Passive and Active
  Measurement (PAM)}}.
\newblock


\bibitem[Partridge and Allman(2016)]%
        {PA16}
\bibfield{author}{\bibinfo{person}{Craig Partridge} {and} \bibinfo{person}{Mark
  Allman}.} \bibinfo{year}{2016}\natexlab{}.
\newblock \showarticletitle{{Addressing Ethical Considerations in Network
  Measurement Papers}}.
\newblock \bibinfo{journal}{\emph{Commun. ACM}} \bibinfo{volume}{59},
  \bibinfo{number}{10} (\bibinfo{date}{Oct.} \bibinfo{year}{2016}).
\newblock


\bibitem[Project(2023)]%
        {zgrab2}
\bibfield{author}{\bibinfo{person}{The~ZMap Project}.}
  \bibinfo{year}{2023}\natexlab{}.
\newblock \bibinfo{booktitle}{\emph{{ZGrab 2.0}}}.
\newblock
\urldef\tempurl%
\url{https://github.com/zmap/zgrab2}
\showURL{%
Retrieved 2023-03-24 from \tempurl}


\bibitem[Sattler et~al\mbox{.}(2023a)]%
        {datatum}
\bibfield{author}{\bibinfo{person}{Patrick Sattler}, \bibinfo{person}{Johannes
  Zirngibl}, \bibinfo{person}{Mattijs Jonker}, \bibinfo{person}{Oliver Gasser},
  \bibinfo{person}{Georg Carle}, {and} \bibinfo{person}{Ralph Holz}.}
  \bibinfo{year}{2023}\natexlab{a}.
\newblock \bibinfo{booktitle}{\emph{{Data and Analysis at TUM University
  Library}}}.
\newblock
\urldef\tempurl%
\url{https://mediatum.ub.tum.de/1723389}
\showURL{%
\tempurl}
\newblock
\shownote{doi:10.14459/2023mp1723389}.


\bibitem[Sattler et~al\mbox{.}(2023b)]%
        {hrpwebsite}
\bibfield{author}{\bibinfo{person}{Patrick Sattler}, \bibinfo{person}{Johannes
  Zirngibl}, \bibinfo{person}{Mattijs Jonker}, \bibinfo{person}{Oliver Gasser},
  \bibinfo{person}{Georg Carle}, {and} \bibinfo{person}{Ralph Holz}.}
  \bibinfo{year}{2023}\natexlab{b}.
\newblock \bibinfo{booktitle}{\emph{{HRP Website with data}}}.
\newblock
\urldef\tempurl%
\url{https://hrp-stats.github.io/}
\showURL{%
Retrieved 2023-10-05 from \tempurl}


\bibitem[Sediqi et~al\mbox{.}(2022)]%
        {sediqi2022hyper}
\bibfield{author}{\bibinfo{person}{Khwaja~Zubair Sediqi}, \bibinfo{person}{Lars
  Prehn}, {and} \bibinfo{person}{Oliver Gasser}.}
  \bibinfo{year}{2022}\natexlab{}.
\newblock \showarticletitle{{Hyper-Specific Prefixes: Gotta Enjoy the Little
  Things in Interdomain Routing}}.
\newblock \bibinfo{journal}{\emph{ACM SIGCOMM Computer Communication Review}}
  \bibinfo{volume}{52} (\bibinfo{date}{June} \bibinfo{year}{2022}).
\newblock
Issue 2.
\urldef\tempurl%
\url{https://doi.org/10.1145/3544912.3544916}
\showDOI{\tempurl}


\bibitem[{Shadowserver}(2023)]%
        {shadowserver}
\bibfield{author}{\bibinfo{person}{{Shadowserver}}.}
  \bibinfo{year}{2023}\natexlab{}.
\newblock \bibinfo{booktitle}{\emph{{Shadowserver - Lighting the way to a more
  secure Internet}}}.
\newblock
\urldef\tempurl%
\url{https://www.shadowserver.org/}
\showURL{%
Retrieved 2023-03-24 from \tempurl}


\bibitem[{Shodan}(2023)]%
        {shodan}
\bibfield{author}{\bibinfo{person}{{Shodan}}.} \bibinfo{year}{2023}\natexlab{}.
\newblock \bibinfo{booktitle}{\emph{{Shodan Dashboard}}}.
\newblock
\urldef\tempurl%
\url{https://www.shodan.io/dashboard}
\showURL{%
Retrieved 2023-03-24 from \tempurl}


\bibitem[Sonar(2023)]%
        {rapid7}
\bibfield{author}{\bibinfo{person}{Rapid7~Project Sonar}.}
  \bibinfo{year}{2023}\natexlab{}.
\newblock \bibinfo{booktitle}{\emph{Open Data}}.
\newblock
\urldef\tempurl%
\url{https://opendata.rapid7.com/}
\showURL{%
Retrieved 2023-03-24 from \tempurl}


\bibitem[van Rijswijk-Deij et~al\mbox{.}(2016)]%
        {openintel}
\bibfield{author}{\bibinfo{person}{Roland van Rijswijk-Deij},
  \bibinfo{person}{Mattijs Jonker}, \bibinfo{person}{Anna Sperotto}, {and}
  \bibinfo{person}{Aiko Pras}.} \bibinfo{year}{2016}\natexlab{}.
\newblock \showarticletitle{A high-performance, scalable infrastructure for
  large-scale active DNS measurements}.
\newblock \bibinfo{journal}{\emph{IEEE journal on selected areas in
  communications}} \bibinfo{volume}{34}, \bibinfo{number}{6}
  (\bibinfo{year}{2016}), \bibinfo{pages}{1877--1888}.
\newblock


\bibitem[Wan et~al\mbox{.}(2020)]%
        {wan2020originofscanning}
\bibfield{author}{\bibinfo{person}{Gerry Wan}, \bibinfo{person}{Liz
  Izhikevich}, \bibinfo{person}{David Adrian}, \bibinfo{person}{Katsunari
  Yoshioka}, \bibinfo{person}{Ralph Holz}, \bibinfo{person}{Christian Rossow},
  {and} \bibinfo{person}{Zakir Durumeric}.} \bibinfo{year}{2020}\natexlab{}.
\newblock \showarticletitle{{On the Origin of Scanning: The Impact of Location
  on Internet-Wide Scans}}. In \bibinfo{booktitle}{\emph{Proceedings of the ACM
  Internet Measurement Conference}} (Virtual Event, USA)
  \emph{(\bibinfo{series}{IMC '20})}. \bibinfo{publisher}{Association for
  Computing Machinery}, \bibinfo{address}{New York, NY, USA},
  \bibinfo{pages}{662–679}.
\newblock
\showISBNx{9781450381383}
\urldef\tempurl%
\url{https://doi.org/10.1145/3419394.3424214}
\showDOI{\tempurl}


\bibitem[Zirngibl et~al\mbox{.}(2022a)]%
        {zirngibl2022prevalenceofparking}
\bibfield{author}{\bibinfo{person}{Johannes Zirngibl}, \bibinfo{person}{Steffen
  Deusch}, \bibinfo{person}{Patrick Sattler}, \bibinfo{person}{Juliane
  Aulbach}, \bibinfo{person}{Georg Carle}, {and} \bibinfo{person}{Mattijs
  Jonker}.} \bibinfo{year}{2022}\natexlab{a}.
\newblock \showarticletitle{{Domain Parking: Largely Present, Rarely
  Considered!}}. In \bibinfo{booktitle}{\emph{Proc. Network Traffic Measurement
  and Analysis Conference (TMA) 2022}}.
\newblock


\bibitem[Zirngibl et~al\mbox{.}(2023)]%
        {zirngibl2023svcb}
\bibfield{author}{\bibinfo{person}{Johannes Zirngibl}, \bibinfo{person}{Patrick
  Sattler}, {and} \bibinfo{person}{Georg Carle}.}
  \bibinfo{year}{2023}\natexlab{}.
\newblock \showarticletitle{{A First Look at SVCB and HTTPS DNS Resource
  Records in the Wild}}. In \bibinfo{booktitle}{\emph{2023 IEEE European
  Symposium on Security and Privacy Workshops (EuroS\&PW)}}.
  \bibinfo{pages}{470--474}.
\newblock


\bibitem[Zirngibl et~al\mbox{.}(2022b)]%
        {zirngibl2022clusters}
\bibfield{author}{\bibinfo{person}{Johannes Zirngibl}, \bibinfo{person}{Lion
  Steger}, \bibinfo{person}{Patrick Sattler}, \bibinfo{person}{Oliver Gasser},
  {and} \bibinfo{person}{Georg Carle}.} \bibinfo{year}{2022}\natexlab{b}.
\newblock \showarticletitle{{Rusty Clusters? Dusting an IPv6 Research
  Foundation}}. In \bibinfo{booktitle}{\emph{Proc. ACM Int. Measurement
  Conference (IMC)}} (Nice, France).
\newblock


\end{thebibliography}
\label{lastpage}

\clearpage
\appendix
\section{Appendix}

\begin{figure}
	\centering
	\begin{subfigure}{0.45\textwidth}
		\includegraphics{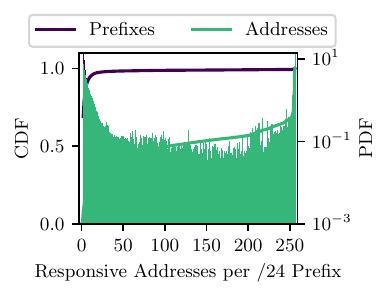}
		\caption{TCP/8443}
		\label{fig:reachable_addrs_8443_per_pfx}
	\end{subfigure}
	\hfill
	\begin{subfigure}{0.45\textwidth}
		\includegraphics{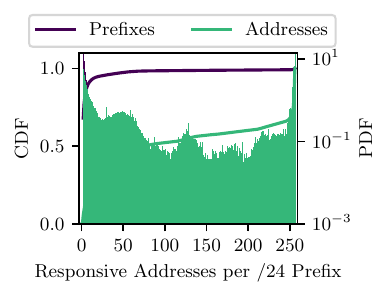}
		\caption{TCP/8080}
		\label{fig:reachable_addrs_8080_per_pfx}
	\end{subfigure}
    \begin{subfigure}{0.45\textwidth}
		\includegraphics{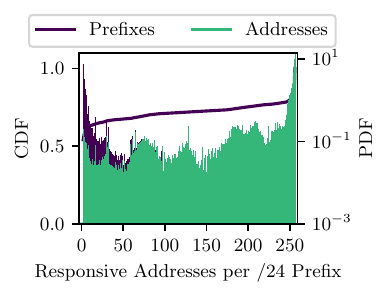}
		\caption{TCP/8984}
		\label{fig:reachable_addrs_8984_per_pfx}
	\end{subfigure}
	\hfill
	\begin{subfigure}{0.45\textwidth}
		\includegraphics{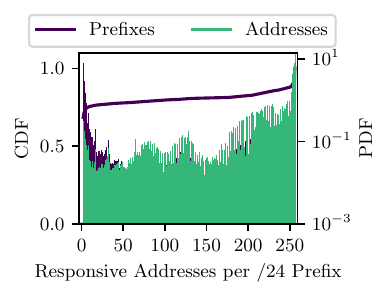}
		\caption{TCP/60000}
		\label{fig:reachable_addrs_60000_per_pfx}
	\end{subfigure}

	\caption{Respective and cumulative probability distribution of responsive addresses inside prefixes. The address data represents the impact of these prefixes on the scan results. Note the logarithmic axis for the right Y-axis.}
	\label{fig:reachable_addrs_8443_8080_per_pfx}
\end{figure}

\Cref{fig:reachable_addrs_8443_8080_per_pfx} shows the distribution of prefix responsiveness for the two most common HTTP/S alternative ports, TCP/60000, and TCP/8984.
They show similar properties to \Cref{fig:reachable_addrs_443_80_per_pfx} and also confirm the joint evaluation in \Cref{fig:reachable_addrs_per_pfx}.
TCP/60000 and TCP/8984 exhibit an even more pronounced bias towards \acp{hrp}.

\received{June 2023}
\received[accepted]{October 2023}

\end{document}